\documentclass[nofootinbib,superscriptaddress,a4paper, twocolumn, floatfix]{revtex4-2}
\usepackage[english]{babel}
\usepackage[utf8]{inputenc}
\usepackage[colorinlistoftodos, color=green!40, prependcaption]{todonotes}
\usepackage{amsthm}
\usepackage{mathtools}
\usepackage{physics}
\usepackage{braket}
\usepackage{listings}
\usepackage{xcolor}
\usepackage{graphicx}
\usepackage[left=23mm,right=13mm,top=35mm,columnsep=15pt]{geometry} 
\usepackage{adjustbox}
\usepackage{placeins}
\usepackage[T1]{fontenc}
\usepackage{lipsum}
\usepackage{csquotes}
\usepackage[pdftex, pdftitle={Article}, pdfauthor={Author}]{hyperref} 
\usepackage{booktabs}
\usepackage{amssymb}
\usepackage{bm}
\usepackage[ruled,vlined]{algorithm2e}
\usepackage[caption=false]{subfig}

\begin{document}

\title{Leveraging commuting groups for an efficient variational Hamiltonian ansatz}

\author{Abhinav Anand}
\email[E-mail:]{abhinav.anand@duke.edu}
\affiliation{Duke Quantum Center, Duke University, Durham, NC 27701, USA.}
\affiliation{Department of Electrical and Computer Engineering, Duke University, Durham, NC 27708, USA.}

\author{Kenneth R. Brown}
\email[E-mail:]{kenneth.r.brown@duke.edu}
\affiliation{Duke Quantum Center, Duke University, Durham, NC 27701, USA.}
\affiliation{Department of Electrical and Computer Engineering, Duke University, Durham, NC 27708, USA.}
\affiliation{Department of Physics, Duke University, Durham, NC 27708, USA.}
\affiliation{Department of Chemistry, Duke University, Durham, NC 27708, USA.}
\date{\today}

\begin{abstract}
    Efficiently calculating the low-lying eigenvalues of Hamiltonians, written as sums of Pauli operators, is a fundamental challenge in quantum computing.  While various methods have been proposed to reduce the complexity of quantum circuits for this task, there remains room for further improvement. In this article, we introduce a new circuit design using commuting groups within the Hamiltonian to further reduce the circuit complexity of Hamiltonian-based quantum circuits. Our approach involves partitioning the Pauli operators into mutually commuting clusters and finding Clifford unitaries that diagonalize each cluster. We then design an ansatz that uses these Clifford unitaries for efficient switching between the clusters, complemented by a layer of parameterized single qubit rotations for each individual cluster. By conducting numerical simulations, we demonstrate the effectiveness of our method in accurately determining the ground state energy of different quantum chemistry Hamiltonians. Our results highlight the applicability and potential of our approach for designing problem-inspired ansatz for various quantum computing applications.
\end{abstract}

\maketitle

\section{Introduction}
The last decade has seen quantum computing emerge as a transformative technology, with the potential to revolutionize various scientific fields~\cite{bharti2022noisy, cerezo2020variational, Anand2021Quantum, mcardle2020quantumreview}.
A critical use of quantum computers involves simulating Hamiltonian time evolution~\cite{feynman1982simulating} for predicting properties of different quantum systems~\cite{aspuru2005simulated}.
However, the existing quantum computing platforms are in their early phases of development and encounter various sources of error, thus restricting the practical applicability of these systems~\cite{preskill2018quantum}.
This requires us to find novel algorithms that are designed to mitigate the effects of noise.
One such method is to design hybrid quantum-classical algorithms~\cite{peruzzo2014variational,farhi2014quantum,McClean2016theoryofvqe} where one utilizes both the classical and quantum computer in a manner that exploits their  respective strengths.

A central object of such algorithms are parameterized quantum circuits (PQCs)~\cite{sim2019expressibility,benedetti2019parameterized,cong2019quantum}, which are used to prepare trial wavefunctions on the quantum computer.
Recent advancements have significantly enhanced our understanding of the design principles~\cite{zhang2022differentiable, du2022quantum}, trainability~\cite{mcclean2018barren, cerezo2021cost, marrero2021entanglement, wang2021noise} convergence properties~\cite{haug2021optimal} and robustness~\cite{weber2022toward} of different PQCs.
A popular approach for design of PQCs are Hamiltonian based circuits~\cite{wecker2015progress} which are known to have better training properties~\cite{wiersema2020exploring} as they preserve the symmetry of the problem.
However, these circuits often possess limitations, such as depth and subspace restrictions, which can impact their effectiveness~\cite{choquette2020quantum,anand2022exploring}.
A potential solution to overcome these limitations was proposed in Ref.~\cite{choquette2020quantum}, where the authors add driving terms to the Hamiltonian to break the problem symmetry and observe better convergence.

Another promising approach involves the utilization of Clifford or near-Clifford circuits for performing useful computation.
These circuits can be simulated classically efficiently~\cite{aaronson2004improved, bravyi2016improved} but are not universal, thus have limited applications.
Nevertheless, they have been used to reduce the number of measurements in quantum algorithms~\cite{verteletskyi2020measurement, izmaylov2019unitary,jena2019pauli, crawford2021efficient, huggins2021efficient,gokhale2020n, zhao2020measurement}, find compressed representation of quantum states~\cite{anand2022quantum}, add correlation to product wavefunctions~\cite{schleich2023partitioning} and for initial state preparation~\cite{ravi2022cafqa}, among others~\cite{anand2025hamiltonian}.

In this study, we use techniques for partitioning of a Hamiltonian into commuting groups and present a novel circuit design that integrates circuits from problem-specific knowledge with general single qubit rotation gates. 
We employ efficient clustering techniques to construct sets of mutually commuting operators and Clifford unitaries, which simultaneously diagonalize these operator sets.
Subsequently, we utilize these Clifford circuits to create ``single-code" and ``combined-codes" ans\"atze, 
where the Clifford circuits define a symmetric subspace of the Hamiltonian and the general rotations navigate these subspaces.
We then apply these circuits to approximate ground state energies of various molecules.
Finally we provide empirical evidence of better convergence of these circuit when compared to the traditional problem-based ansatz.

The remaining sections of this paper are organized as follows: Section~\ref{sec:method} outlines the preliminary information and the method used in this study. 
Section~\ref{sec:simulations} presents the results from numerical simulations, and finally, Section~\ref{sec:conclusion} provides concluding remarks.

\section{Methodology}\label{sec:method}

\subsection{Clustering Hamiltonian into commuting groups}\label{subsec:Ham_group}
A quantum Hamiltonian, $\hat{H}$, can be written as
\begin{equation}\label{eq:qubit_ham}
\hat{H} = \sum_{k=1}^{M} c_k \hat{P}_k,
\end{equation}
where $c_k$ is a complex number and $\hat{P}_k$ is a Pauli-string on $n$ qubits.
A Pauli-string is defined as the tensor product of Pauli matrices ($\hat{\sigma}_x, \hat{\sigma}_y, \hat{\sigma}_z$) and the identity operator $\hat{I}$ as
\begin{equation} 
\hat{P}_k = \bigotimes_{j=1}^{n} \hat{\sigma}, \end{equation}
with $\hat{\sigma} \in \{ \hat{I}, \hat{\sigma}_x, \hat{\sigma}_y, \hat{\sigma}_z \}$.
The Hamiltonian can be further divided into $m$ sets of mutually commuting groups as
\begin{equation}\label{eq:ham_comm}
    \hat{H} = \sum_{k=1}^{m} \sum_{l=1}^{m_i} c_{kl} \hat{P}_{kl};
\end{equation}
\begin{equation}
    [\hat{P}_{ki}, \hat{P}_{kj}] = 0 \text{   } \forall \text{   } (\hat{P}_{ki}, \hat{P}_{kj}) \in \{\hat{P}_{k1}, ... , \hat{P}_{km_i}\} 
\end{equation}
where $\hat{P}_{kl}$ is the $l$-th Pauli-string in the $k$-th commuting set and $c_{kl}$ is the complex coefficient.
It is known~\cite{jena2019pauli} that, given a set of commuting terms, there exists a Clifford circuit $\mathcal{U}$ that simultaneously diagonalizes each operator in the set as
\begin{equation} \label{eq:cliff_diag}
\mathcal{U} \hat{P}_{kl} \mathcal{U}^{\dagger} = \bigotimes_{j=1}^{n} \hat{\sigma}_j, \text{ } \forall \text{ } \hat{P}_{kl} \in {\hat{P}_{k1}, \dots, \hat{P}_{km_i}}, 
\end{equation}
where $\hat{\sigma}_j \in \{ \hat{I}, \hat{\sigma}_z \}$.
In recent years, several proposals~\cite{verteletskyi2020measurement, izmaylov2019unitary, jena2019pauli, crawford2021efficient, huggins2021efficient, gokhale2020n, zhao2020measurement} have been put forward for partitioning the Hamiltonian into sets of commuting groups.
In this work, we follow the techniques presented in Refs.~\cite{crawford2021efficient}.

The gate complexity of the unitary $\mathcal{U}$ depends on the type of commutativity chosen: qubit-wise commutativity versus general commutativity.
In this work, we use the general commutativity approach, as it leads to a smaller number of commuting sets but results in deeper unitaries.

\subsection{Variational Hamiltonian ansatz (VHA)}\label{subsec:VHA}
Given a Hamiltonian, $\hat{H}$, as in Eq.~\ref{eq:qubit_ham} we define an ansatz as
\begin{align}\label{eq:t-VHA}
    U(\boldsymbol{\theta}) &= e^{-i\boldsymbol{\theta}\hat{H}} = e^{-i\sum_{k=1}^{M} \theta_k \hat{P}_k} \nonumber \\
        &\approx \prod_{k=1}^{M} e^{-i \theta_k \hat{P}_k} 
\end{align}
where $\{ \theta_k \}$ are variational parameters.
Here, we have used the first order Trotter-Suzuki approximation~\cite{suzuki1976generalized} to decompose the exponential map of the Hamiltonian, $e^{-i\boldsymbol{\theta}\hat{H}}$, into products of exponential maps of Pauli-strings, $e^{-i \theta_k \hat{P}_k}$.
The resultant unitary is the variational Hamiltonian ansatz~\cite{wecker2015progress}.

\begin{figure}[htbp!]
\centering
\subfloat[A schematic of the traditional  variational Hamiltonian ansatz (VHA) of the form in Eq.~\ref{eq:t-VHA}.  A blue box represents a gate of the form $e^{-i\theta\hat{P}_i}$, where $\hat{P}_i$ can be any Pauli-string.]{\includegraphics[width=0.45\textwidth]{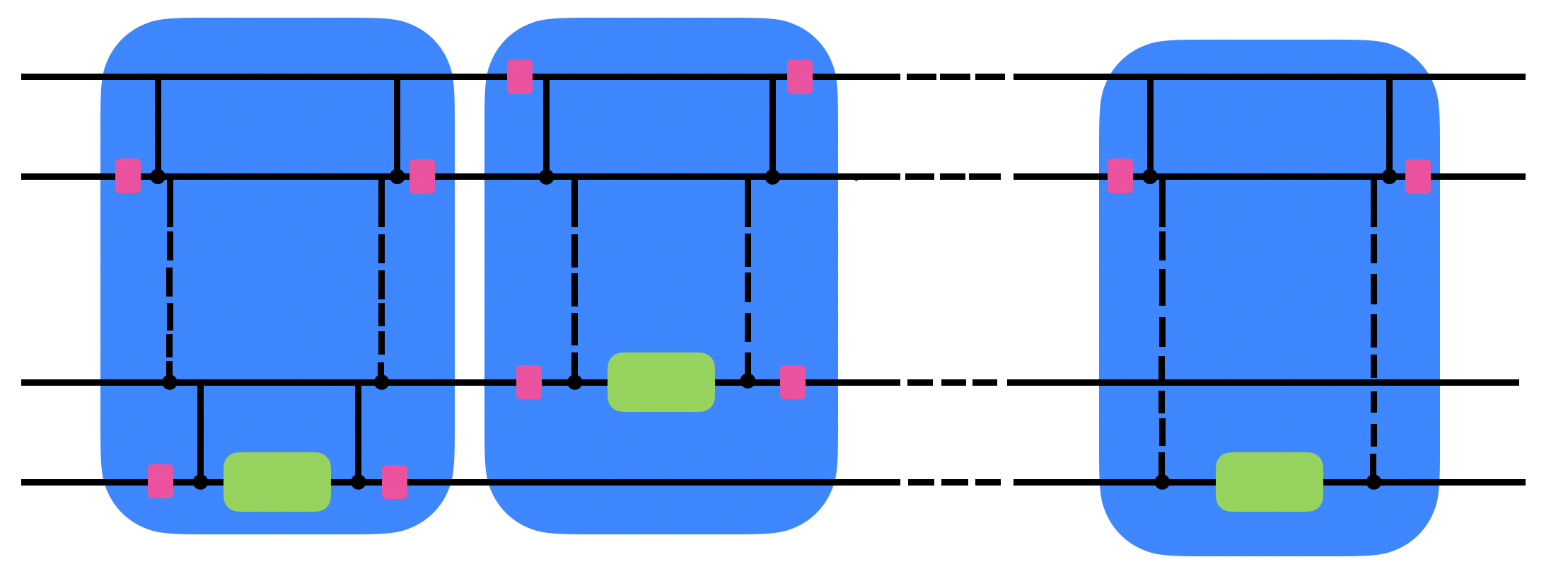}} \\
\subfloat[A schematic of the variational Hamiltonian ansatz of the form in Eq.~\ref{eq:VHA_}. A blue box represents a gate of the form $e^{-i\theta\hat{P}_i}$, where $\hat{P}_i$'s are diagonal Pauli-strings.]{\includegraphics[width=0.45\textwidth]{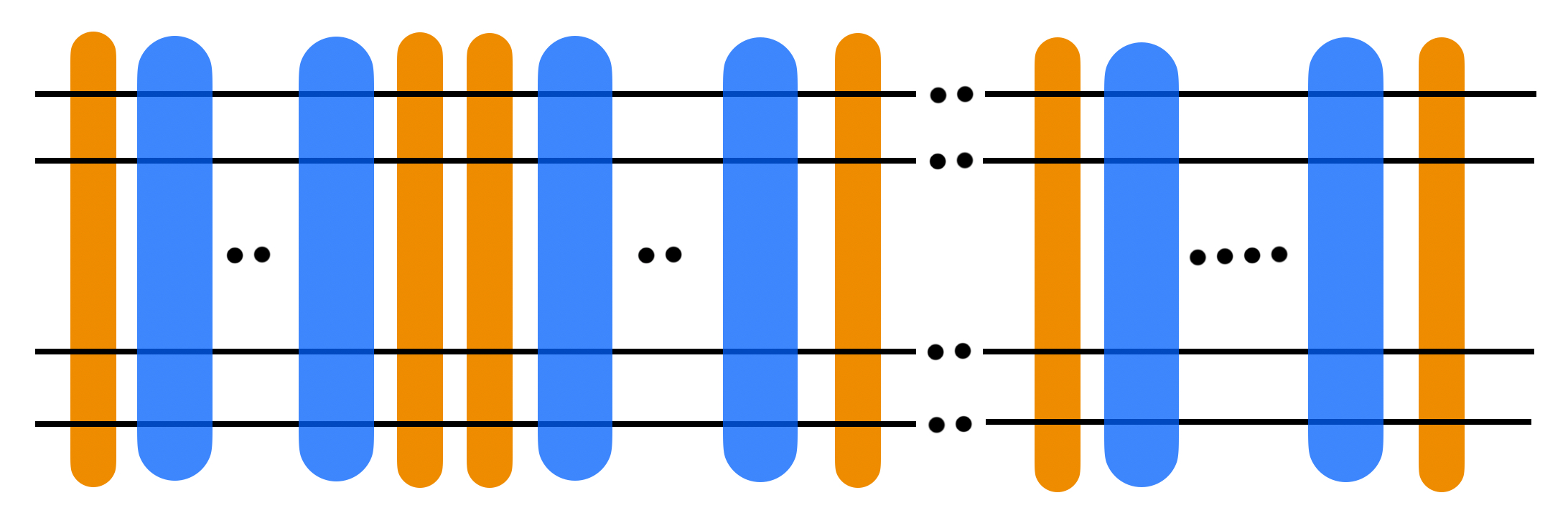}} \\
\caption{\label{fig:VHA}A schematic of the different forms of the variational Hamiltonian ansatz.  The green boxes represent gate of the form $e^{-i\theta\hat{\sigma}_z}$, the pink boxes represents gates for basis change and the orange boxes represents Clifford circuits ($\mathcal{U}_i$ and $\mathcal{U}_i^{\dagger}$) for simultaneous diagonalization.} 
\end{figure}

Furthermore, by partitioning the Hamiltonian into commuting groups (Eq.~\ref{eq:ham_comm}) and using the Clifford circuits in Eq.~\ref{eq:cliff_diag}, we can further write the unitary as
\begin{equation}\label{eq:VHA_}
    U(\boldsymbol{\theta}) = \prod_{k=1}^{m} \mathcal{U}^{\dagger}_k (\prod_{l=1}^{m_k} e^{-i\theta_{kl}\hat{P}_{kl}}) \mathcal{U}_k,
\end{equation}
where $\{\theta_{kl}\}$ are variational parameters.
The variational Hamiltonian ansatz is a product of unitaries that correspond to short time evolution under different parts of the Hamiltonian and can be repeated multiple times to get better approximation of the full time evolution unitary, $e^{-i\boldsymbol{\theta}\hat{H}}$.
A schematic representation of the VHA circuit is shown in Fig.~\ref{fig:VHA}.
This ansatz has been used for approximating eigenvalues of different condensed matter systems as well as for strongly correlated systems in quantum chemistry.
However, they have been known to have some issues~\cite{wiersema2020exploring, anand2022exploring} such as, limited expressibility, larger circuit depths, among others.
In what follows, we present our proposed method that modifies the VHA circuit to mitigate some of the issues.

\subsection{Modified Variational Hamiltonian Ansatz}\label{subsec:M-VHA}
Given a decomposition of the Hamiltonian as in Eq.~\ref{eq:ham_comm}, each set of mutually commuting terms $\{\hat{P}_{k1}, ... , \hat{P}_{km_i}\}$ forms an abelian group.
We can then construct a stabilizer group~\cite{gottesman1997stabilizer} corresponding to each commuting group by replacing some of the Pauli operators, $\hat{P}_{ki}$ with -$\hat{P}_{ki}$.
The full procedure for constructing these stabilizer groups is as follows:
\begin{enumerate}
    \item Collect all the terms $\hat{P}_{k}$ in the Hamiltonian, $\hat{H} = \sum_{k=1}^{M} c_k \hat{P}_k$, which are tensor products of only $\hat{\sigma}_z$ and $\hat{I}$, in one set.
    \item Use the technique in Ref.\cite{crawford2021efficient} to determine the remaining sets and the unitaries required for simultaneous diagonalization. At this stage, we have $m$ sets ${ \mathcal{G}_i }$ and $m$ unitaries ${ \mathcal{U}_i }$ that diagonalize the operators within each set.
    \item The stabilizer group $\mathcal{S}_i$ corresponding to the set with only $\hat{\sigma}_z$ and $\hat{I}$ can be constructed by replacing $\hat{P}_{ki}$ with $-\hat{P}_{ki}$ if there is an odd number of $\hat{\sigma}_z$ operators acting on the first $n_e$ qubits, where $n_e$ is the number of electrons. A state $\ket{\Psi{s_i}}$, stabilized by this group, is the Hartree-Fock state, $\ket{\text{HF}}$.
    \item For the remaining groups, we use their corresponding diagonal representations (which can be obtained using a similarity transform with the unitaries determined above) to replace $\hat{P}_{ki}$ with $-\hat{P}_{ki}$, following the same procedure as in Step 3. A stabilizer state for these groups is given by $\ket{\Psi_{s_i}} = \mathcal{U}_i^{\dagger} \ket{\text{HF}}$.
\end{enumerate}

The stabilizer groups ${\mathcal{S}_i}$ constructed above can be regarded as error-detecting codes, where the elements of the group are the stabilizers and the states ${\ket{\Psi{s_i}}}$ define the codespace.
We can use these stabilizer states ${\ket{\Psi_{s_i}}}$ and the unitaries $\mathcal{U}_i$ to construct a modified VHA ansatz.
We describe the construction in detail in the following sections.

\subsubsection{Single-code ansatz}\label{subsubsec:SCA}
For every group $\mathcal{G}_i$, we construct a Hamiltonian $\hat{H}^{'} = \sum_{g_j \in  \mathcal{G}_i } g_{j}$, which is the sum of all the operators in the group $\mathcal{G}_i$.
This Hamiltonian, after a similarity transform under the Clifford unitary $\mathcal{U}_i$, becomes a diagonal operator.
Consequently, the eigenvectors of the transformed Hamiltonian correspond to computational basis states.

\begin{figure}[htbp!] \centering \includegraphics[width=0.225\textwidth]{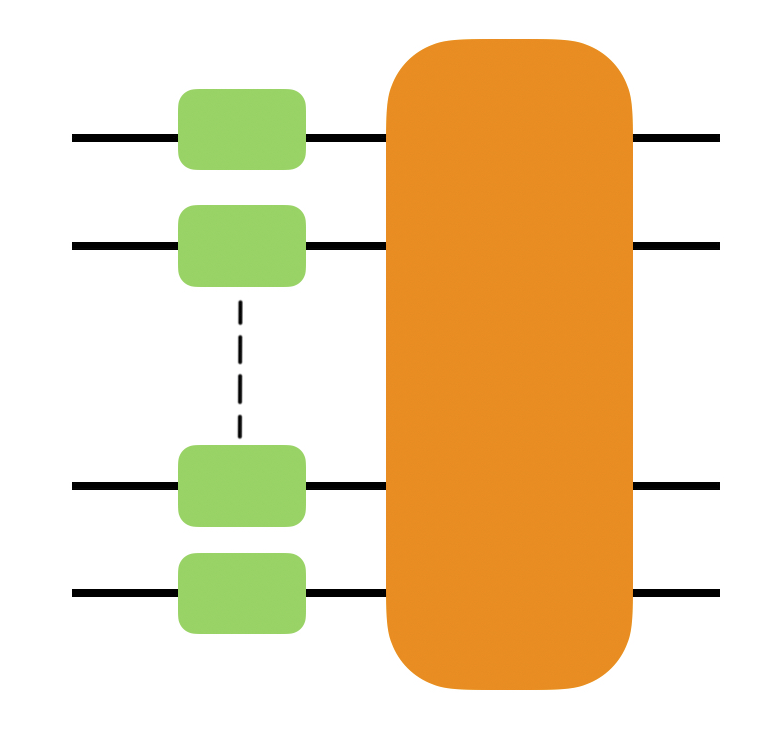} \caption{\label{fig:SCA} A schematic of a single layer of the single-code ansatz. The orange box represents Clifford circuits ($\mathcal{U}_i^{\dagger}$), and the green boxes represent general single-qubit rotation gates of the form $e^{-i\theta_x \hat{\sigma}_x} e^{-i\theta_y \hat{\sigma}_y} e^{-i\theta_z \hat{\sigma}_z}$.} 
\end{figure}

To construct the ground state of the Hamiltonian $\hat{H}^{'}$, we modify the stabilizer states for each group, $\ket{\Psi_{s_i}} = \mathcal{U}_i^{\dagger} \ket{\text{HF}}$, by adding a layer of general single-qubit rotations to create a near-Clifford state.
The resulting ansatz takes the form:
\begin{equation} 
U_{s_{i}}(\boldsymbol{\theta}_i) = \mathcal{U}_i^{\dagger} \left(\bigotimes_{j=1}^{n} \textbf{Rx}_{j}(\theta{x_j})\textbf{Ry}_{j}(\theta{y_j})\textbf{Rz}_{j}(\theta{z_j})\right), 
\end{equation} 
where $\mathcal{U}_{i}$ is the unitary that diagonalizes all the operators in the group $\mathcal{G}_i$, $\textbf{Rx}$, $\textbf{Ry}$, and $\textbf{Rz}$ are single-qubit rotation gates, and $\theta_{x_j}$, $\theta_{y_j}$, and $\theta_{z_j}$ are variational parameters.
We refer to an ansatz of this form as the single-code ansatz (Fig.~\ref{fig:SCA}), as it is derived from a single Abelian group.

This ansatz can be used to find the ground state of the Hamiltonian $\hat{H}^{'}$ by minimizing the following objective function:
\begin{equation} 
E(\boldsymbol{\theta}) = \bra{\text{HF}} U_{s{i}}^{\dagger}(\boldsymbol{\theta}) \hat{H}^{'} U_{s{i}}(\boldsymbol{\theta})\ket{\text{HF}}. 
\end{equation} 
This optimization corresponds to finding the ground state of a block of the Hamiltonian $\hat{H}^{'}$.
The resulting state $U_{s{_i}}(\boldsymbol{\theta}^{*})\ket{\text{HF}}$ can be regarded as an entangled mean-field solution, a concept that has also been explored in previous studies as well~\cite{ryabinkin2018constrained}.
We hypothesize that the state $U_{s{_i}}(\boldsymbol{\theta}^{*})\ket{\text{HF}}$, which is a classically simulatable state, can serve as a better initial state than the Hartree-Fock state for the full Hamiltonian.

A classically simulatable state here refers to one for which expectation values of Pauli observables can be efficiently evaluated on a classical computer. 
This is possible because the circuit preparing the state consists of a layer of single-qubit non-Clifford gates followed by a layer of Clifford gates, which transforms a Pauli observable into another Pauli operator. 
As a result, the task reduces to calculating the expectation value with respect to a product state, which is classically efficient.

We provide empirical evidence to support this hypothesis through numerical simulations, and we report the results in Sec.~\ref{sec:simulations}.

\subsubsection{Combined-codes ansatz}\label{subsubsec:CCA}
\begin{figure}[htbp!]
    \centering
    \includegraphics[width=0.45\textwidth]{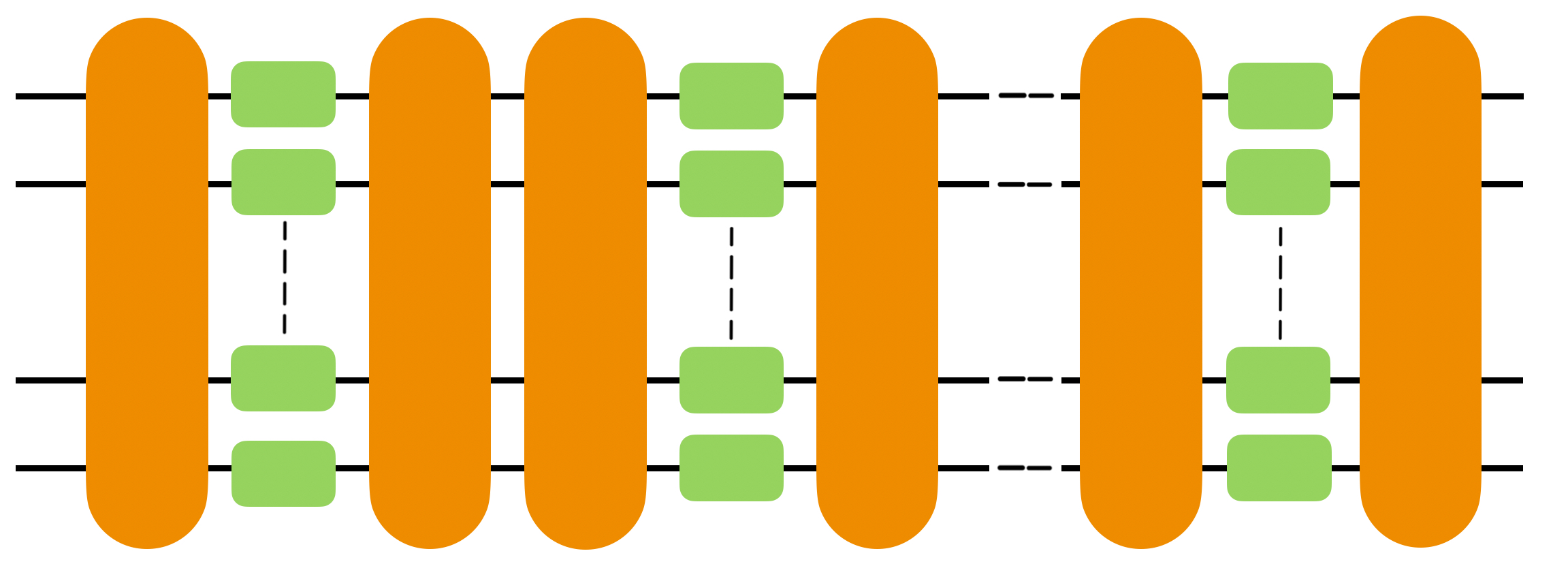}
    \caption{\label{fig:CCA} A schematic of a single layer of the combined-codes ansatz. The orange boxes represent Clifford circuits ($\mathcal{U}_i$ and $\mathcal{U}_i^{\dagger}$) and the green boxes represents a general single qubit rotation gate of the kind $e^{-i\theta_x \hat{\sigma}_x}  e^{-i\theta_y \hat{\sigma}_y} e^{-i\theta_z \hat{\sigma}_z} $.} 
\end{figure}

We can combine all the single-code ans\"atze to form a more complete ansatz by using the unitaries $\mathcal{U}_i\mathcal{U}_j^{\dagger}$ to transition between the bases of different groups $\mathcal{G}_i$ to $\mathcal{G}_j$. 
This combined ansatz, referred to as the combined-codes ansatz(Fig.~\ref{fig:CCA}) can be expressed as:
\begin{align}
    U(\boldsymbol{\theta}) = \prod_{i}^{m} \mathcal{U}_i^{\dagger} \bigotimes_{j=1}^{n} \textbf{Rx}_{j}(\theta_{x_{i,j}})\textbf{Ry}_{j}(\theta_{y_{i,j}})\textbf{Rz}_{j}(\theta_{z_{i,j}})  \mathcal{U}_i,
\end{align}
where $\theta_{x_{i,j}}$, $\theta_{y_{i,j}}$ and $\theta_{z_{i,j}}$ are variational parameters. 

The ordering of the single-code ansätze within the combined ansatz plays a critical role in ensuring better convergence properties.
To determine this ordering, adaptive strategies~\cite{tang2021qubit, grimsley2019adaptive, ryabinkin2018qubit, ryabinkin2020iterative} can be used to sequentially combine the single-code ansätze and construct the full ansatz.
However, such adaptive strategies may introduce additional overhead due to the need for extra circuit evaluations.
Instead, we employ a simpler strategy based on the 1-norm of the group to determine the ordering.

The proposed construction results in shorter circuits~\cite{van2020circuit} with improved convergence properties, as it introduces more degrees of freedom by incorporating general single-qubit unitaries within the subspace spanned by each commuting group.
Furthermore, the total number of parameters in the circuit scales linearly with both the number of qubits, $n$, and the number of commuting groups, $m$.

We hypothesize that this ansatz can effectively approximate low-energy eigenstates and eigenenergies within the Variational Quantum Eigensolver (VQE) framework.
To validate this hypothesis, we conduct numerical experiments on various molecular systems and present our findings in Sec.~\ref{sec:simulations}.
The accuracy of the approximation can be further improved by repeating the combined-codes ansatz multiple times.

As an illustration of the proposed framework, we provide a detailed construction of the different ansätze for the hydrogen molecule in Appendix~\ref{sec:appendix}.

\section{Numerical Experiments}\label{sec:simulations}
In the following we will illustrate some applications of the proposed ansatz. 
We have implemented the whole framework using the Tequila~\cite{kottmann2021tequila} package, which uses Qulacs~\cite{suzuki2021qulacs} as the backend for executing numerical simulations.
We utilize the BFGS algorithm provided by SciPy~\cite{virtanen2020scipy} for gradient-based optimization.

To map fermionic Hamiltonians of various molecules to qubit Hamiltonians of the form given in Eq.\ref{eq:qubit_ham}, we employ the Jordan-Wigner transformation\cite{jordan1928pauli}.
The Hamiltonians used in the numerical simulations can be accessed here~\cite{github_Ham_MVHA}.

For all the numerical simulations, the initial values of the circuit parameters were set to 0.001 and we set the convergence criteria to either a maximum of 100 iterations or a step size of $10^{-6}$.
All energy values are in Hartree (Ha) units and all bond length values are in Angstrom (\AA) units,  unless specified otherwise.
Furthermore, we note that all simulations are ideal and do not include hardware or shot noise.

\subsection{Simulations with single layer}\label{subsec:singlelayer}
Here we present results from numerical simulations using the single-code ansatze and a layer of the combined-codes ansatz for approximating ground state energies of different molecules using the VQE framework. 

\subsubsection{Small Molecules: H$_{2}$ and LiH}\label{subsubsec:H2andLiH}
\begin{figure}[htbp!]
    \centering
    \begin{tabular}{c c}
    \toprule
    \textbf{a) H$_{2}$ molecule}   & \textbf{b)  LiH molecule }\\
    \midrule
    \multicolumn{2}{c}{\textbf{Energy}}\\
    \midrule
    \includegraphics[width=0.45\columnwidth]{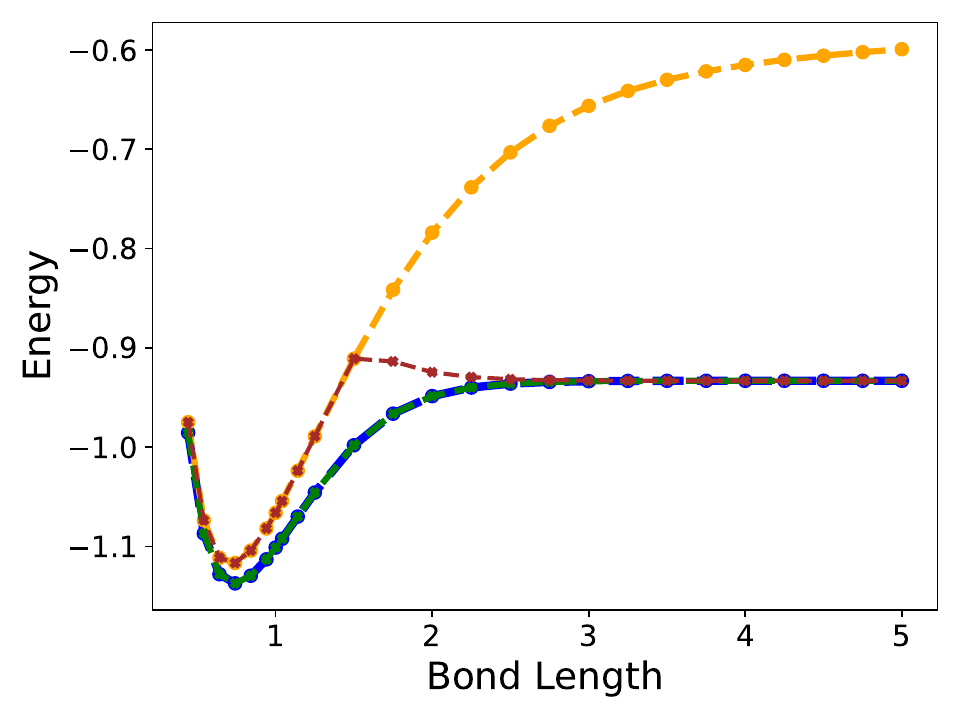} & 
    \includegraphics[width=0.45\columnwidth]{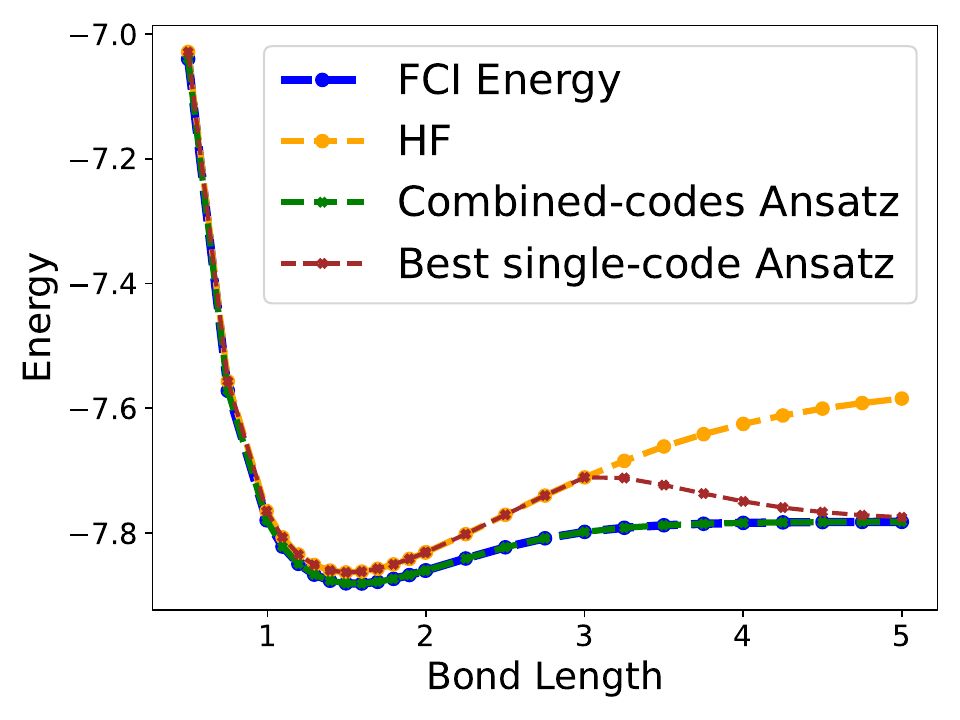}\\
    \midrule
    \multicolumn{2}{c}{\textbf{Error in energy}}\\
    \midrule
    \includegraphics[width=0.45\columnwidth]{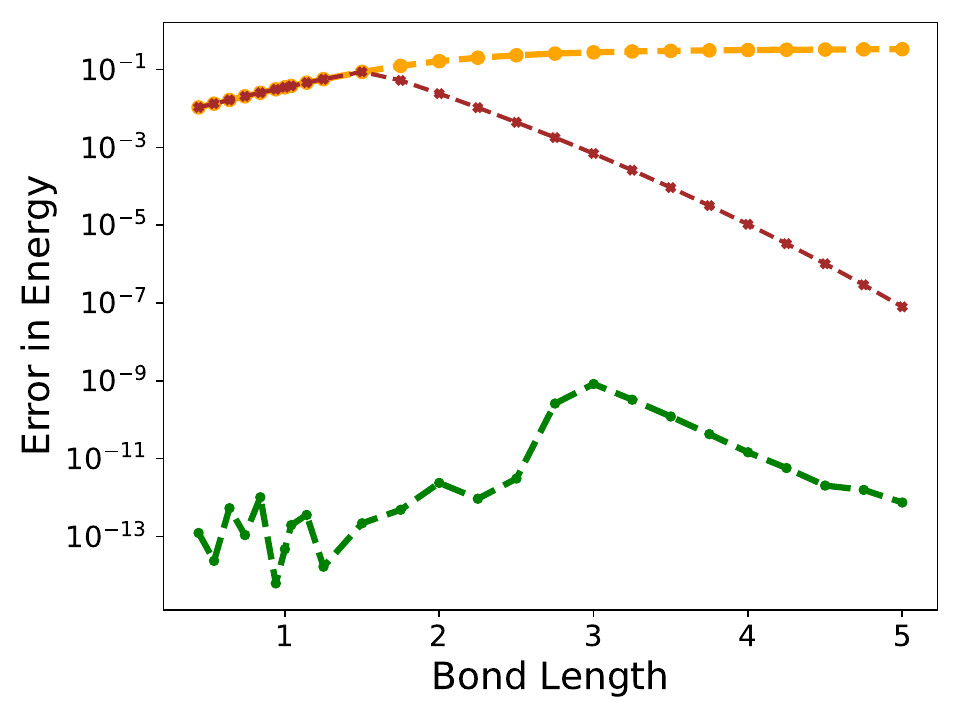} & 
    \includegraphics[width=0.45\columnwidth]{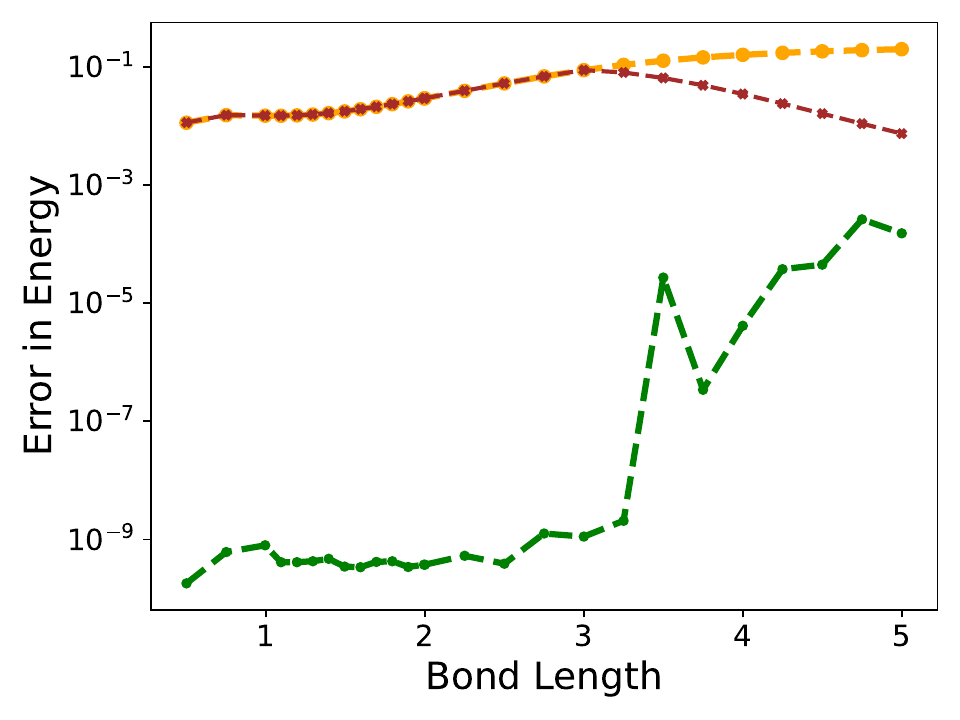}\\
    \midrule
    \end{tabular}
    \caption{\label{fig:H2andLiH} VQE results from optimization of ground state energies for H$_{2}$ and LiH molecule using different geometries with a single layer of the proposed ansatz.} 
\end{figure}

We first apply our ansatz to approximate the ground state energies of the hydrogen molecule (H$_2$) in the minimal basis, consisting of two electrons in four spin-orbitals, and the lithium hydride molecule (LiH) in an active space of the minimal basis, with two electrons in six spin-orbitals. 
The results of the simulations are presented in Fig.~\ref{fig:H2andLiH}.

From Fig.~\ref{fig:H2andLiH}, we observe that for both H$_2$ and LiH, the combined-codes ansatz converges to the exact solution (FCI) within chemical accuracy (10$^{-3}$ Hartree) for all geometries considered.
This indicates that even a single layer of the proposed ansatz is expressive enough to provide accurate ground-state energy approximations for small molecules.

We also analyze the performance of the single-code ans\"atze by plotting the energy corresponding to the best-performing single-code ansatz (i.e., the ansatz yielding the lowest energy). 
For both H$_2$ and LiH, the single-code ansatz outperforms the Hartree-Fock (HF) energy, particularly for stretched geometries, while converging to the HF energy near equilibrium geometries.

These encouraging results for small molecules motivated us to test the performance of our ansatz on larger systems.

\subsubsection{Larger Molecules: H$_4$, BeH$_2$, H$_{2}$O and N$_{2}$}\label{subsubsec:H2OandN2}
\begin{figure}[htbp!]
    \centering
    \begin{tabular}{c c}
    \toprule
    \textbf{Energy} & \textbf{Error in energy}  \\
    \midrule
    \multicolumn{2}{c}{\textbf{a) H$_4$ molecule}}\\
    \midrule
    \includegraphics[width=0.425\columnwidth]{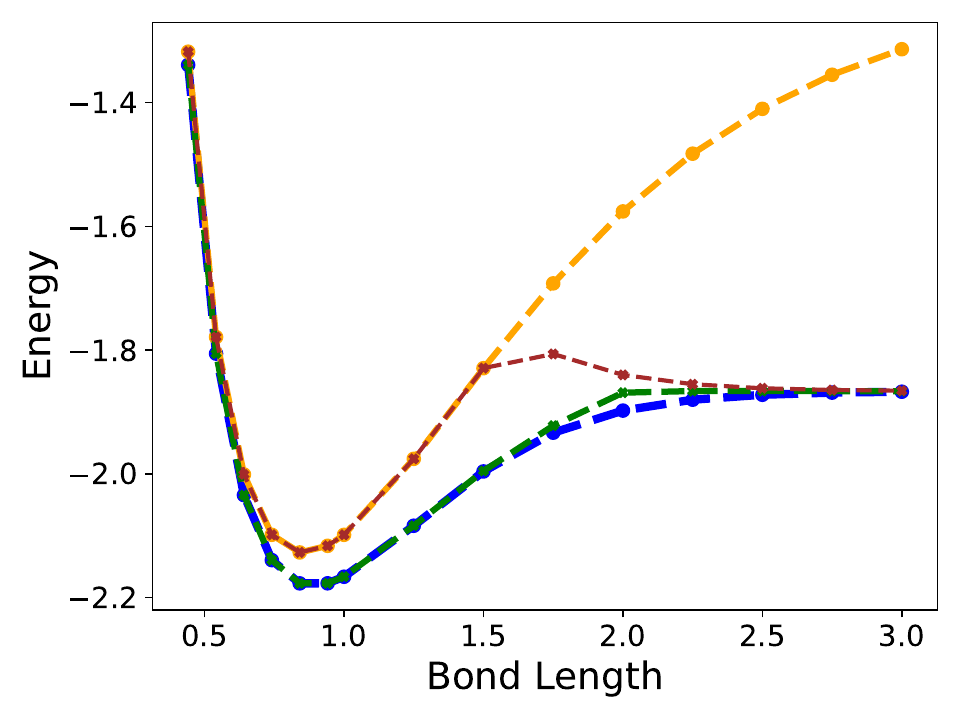} & 
    \includegraphics[width=0.425\columnwidth]{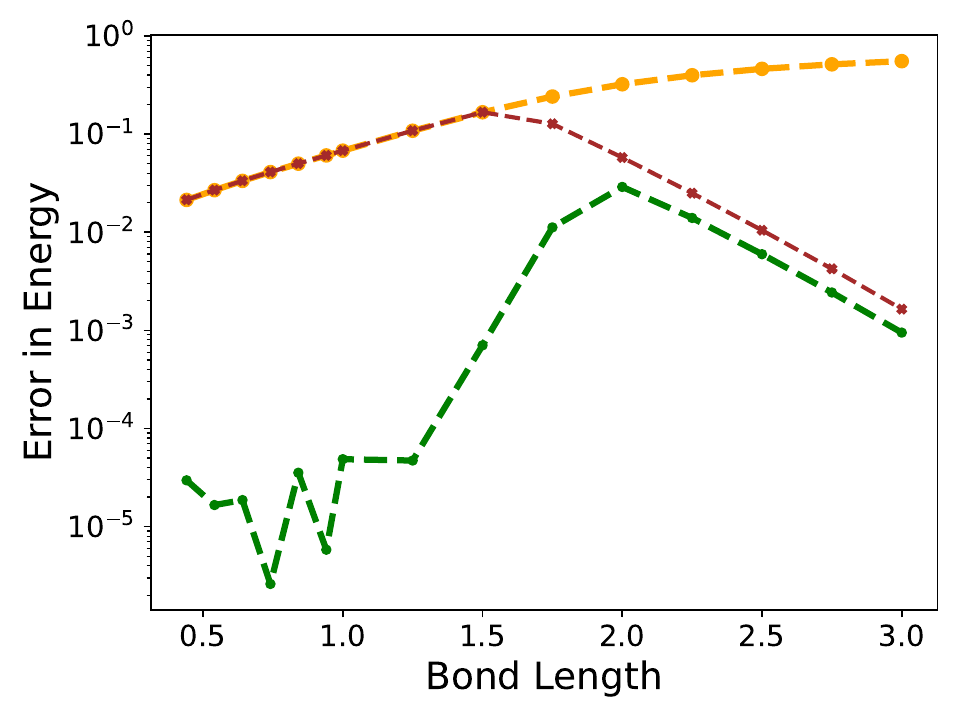} \\
    \midrule
    \multicolumn{2}{c}{\textbf{b) BeH$_2$ molecule}}\\
    \midrule
    \includegraphics[width=0.45\columnwidth]{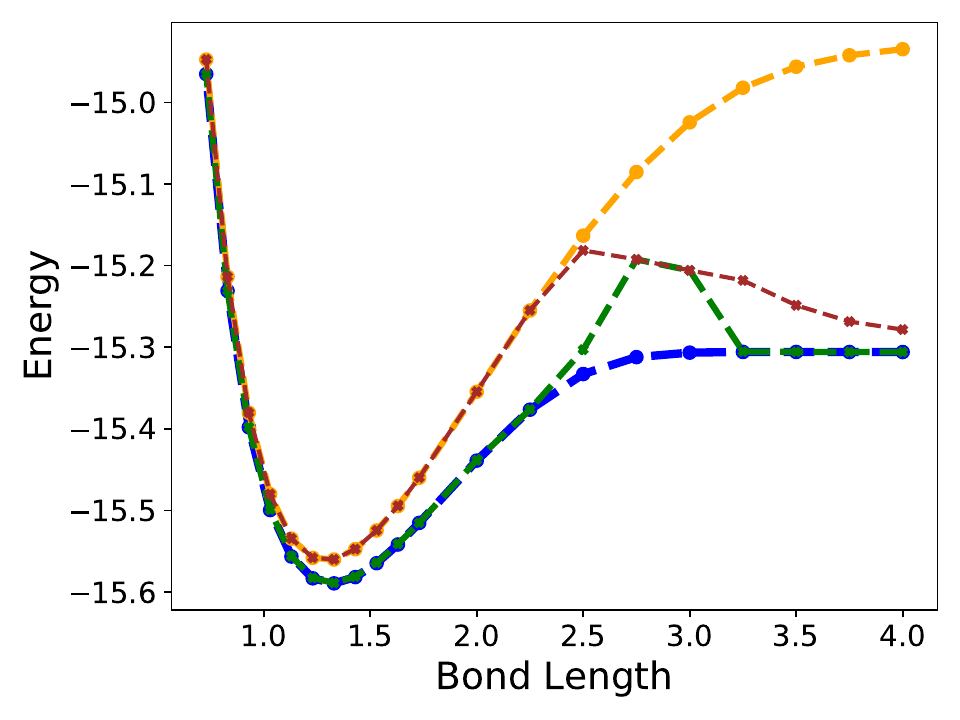} & 
    \includegraphics[width=0.45\columnwidth]{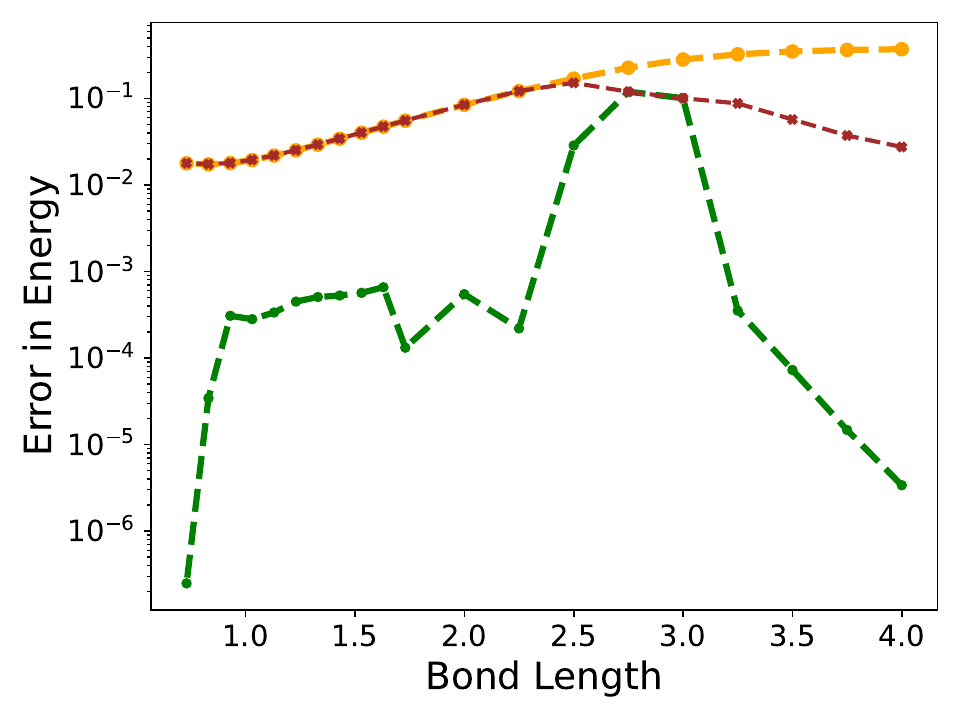} \\
    \midrule
    \multicolumn{2}{c}{\textbf{c) H$_{2}$O molecule}}\\
    \midrule
     \includegraphics[width=0.45\columnwidth]{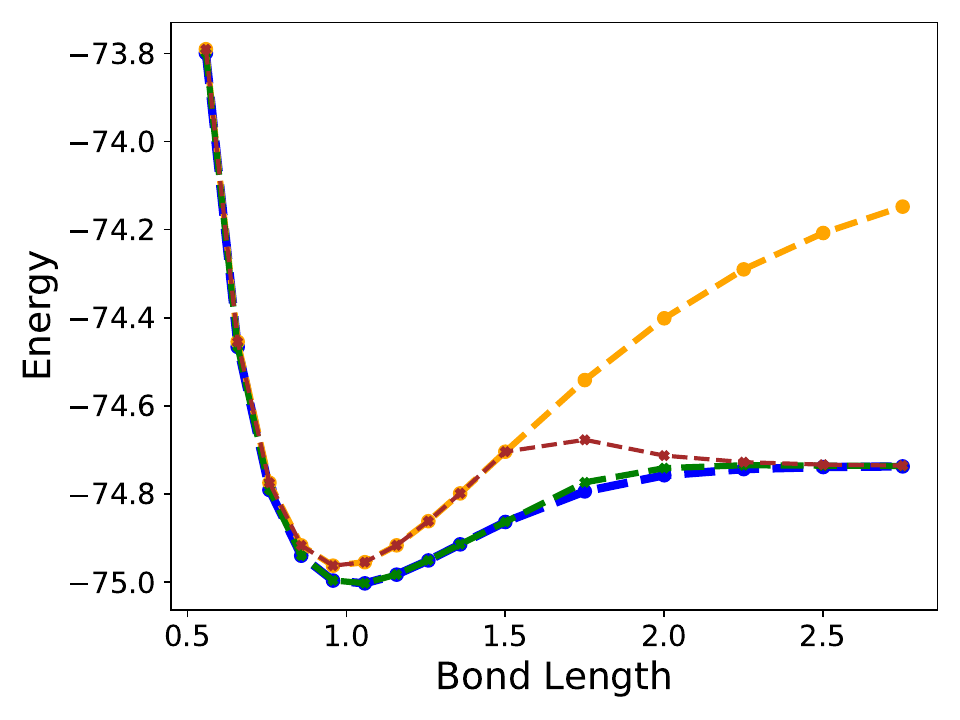} &  
    \includegraphics[width=0.45\columnwidth]{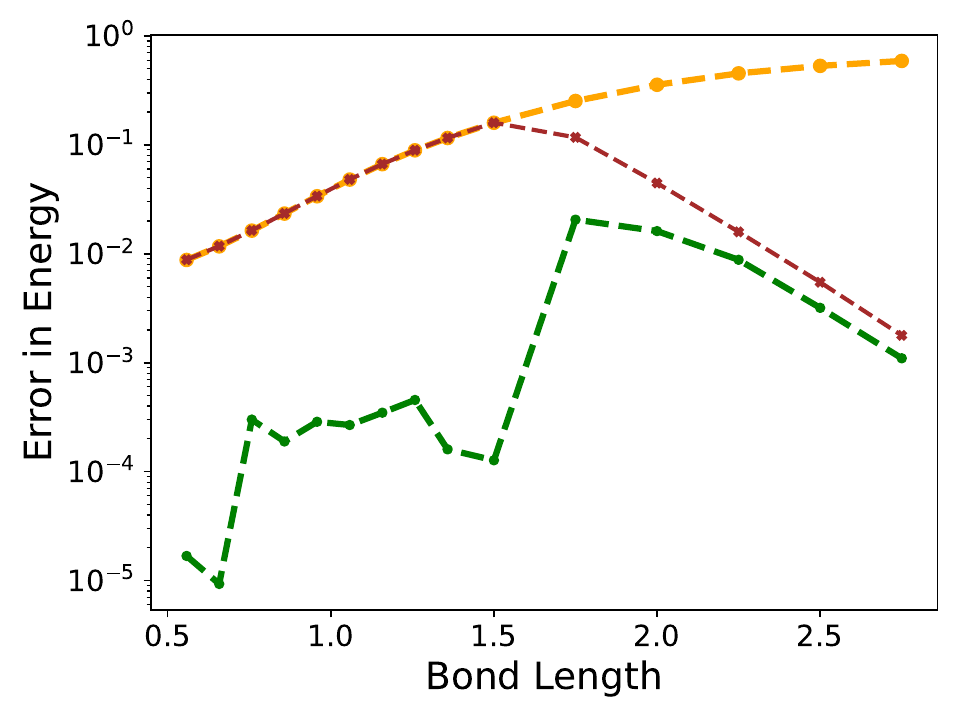} \\ 
    \midrule
    \multicolumn{2}{c}{\textbf{d) N$_{2}$ molecule}}\\
    \midrule
    \includegraphics[width=0.45\columnwidth]{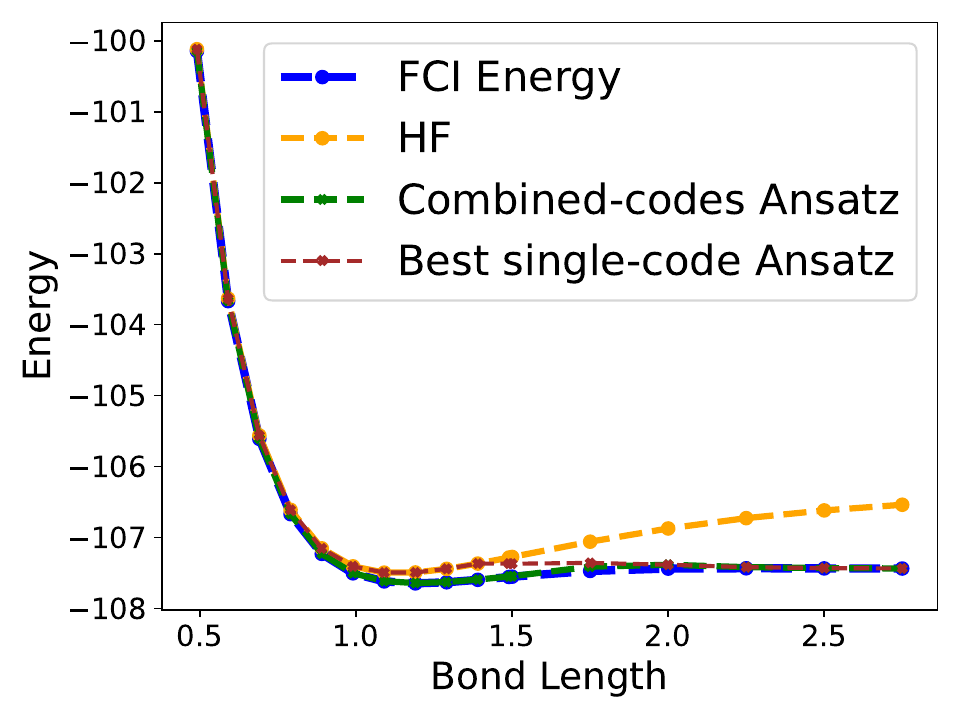} &
    \includegraphics[width=0.45\columnwidth]{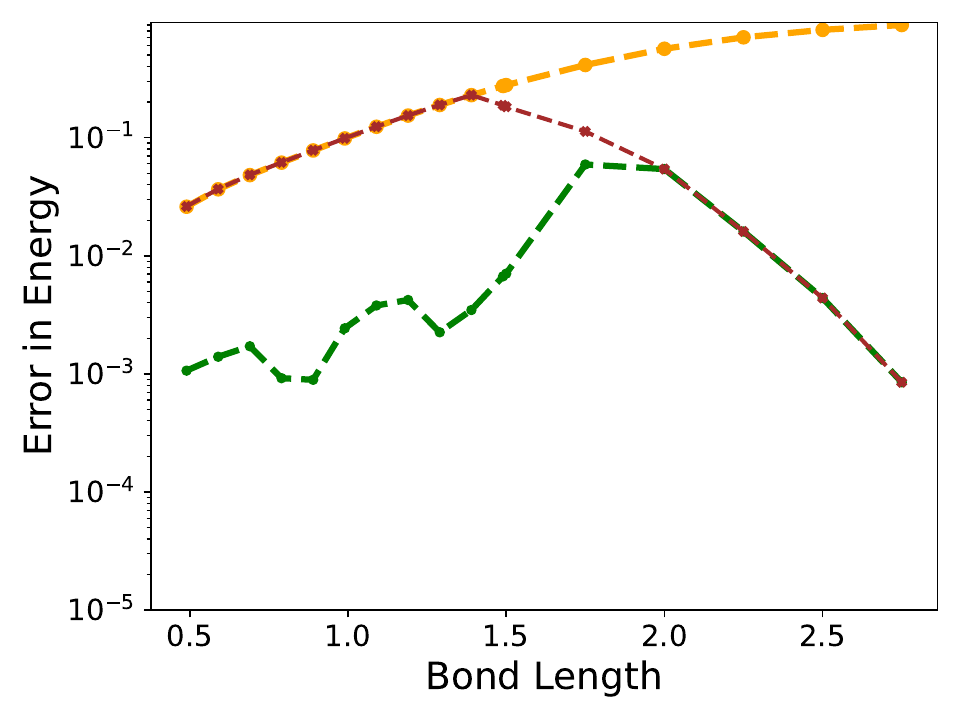}\\
    \midrule
    \end{tabular}
    \caption{\label{fig:H2OandN2} VQE results from optimization of ground state energies for H$_4$, BeH$_2$, H$_{2}$O and N$_{2}$ molecule using different geometries with a single layer of the proposed ansatz.} 
\end{figure}

In this section, we extend our experiments to slightly larger molecules.  
We simulate the linear hydrogen chain (H$_4$), consisting of four electrons in eight spin-orbitals; the active space of the beryllium hydride molecule (BeH$_2$), consisting of four electrons in eight spin-orbitals; the active space of the water molecule (H$_2$O), which has six electrons in ten spin-orbitals; and the active space of the nitrogen molecule (N$_2$), which has six electrons in twelve spin-orbitals.
In the cases of the H$_4$, BeH$_2$ and H$_2$O molecule, we investigate the symmetric stretching of the three H-H, two Be-H and two O-H bonds, respectively. 
The results from all the simulations are shown in Fig.~\ref{fig:H2OandN2}.

The results for these larger molecules mirror those observed for smaller systems with single-code ans\"atze. Specifically, the single-code ansatz converges to the HF energy for geometries near equilibrium but outperforms HF for stretched geometries, suggesting its potential utility for initial state preparation in regimes of strong electronic correlation.

However, the combined-codes ansatz performs less effectively for larger systems compared to smaller ones.
For all the molecules, the optimized energy for configurations near equilibrium geometry are close to chemical accuracy ($10^{-3}$Hartree) when compared with the exact ground-state energy. 
However, for stretched configurations, the ansatz fails to converge to the exact solution.

We attribute this behavior to the limited expressivity of a single layer of the proposed ansatz. 
Similar trends~\cite{anand2025hamiltonian} are observed with Hamiltonian-based ans\"atze, which often require multiple layers to achieve improved approximations.

To address the observed limitations, we now analyze the convergence properties of the proposed ansatz as a function of the number of layers. 
The results of this analysis are reported in the following section.

\subsection{Simulation with two layers}\label{subsec:twolayers}
\begin{figure}[htbp!]
    \centering
    \begin{tabular}{c}
    \toprule\textbf{a) H$_4$ molecule } \\
    \midrule
    \includegraphics[width=0.35\textwidth]{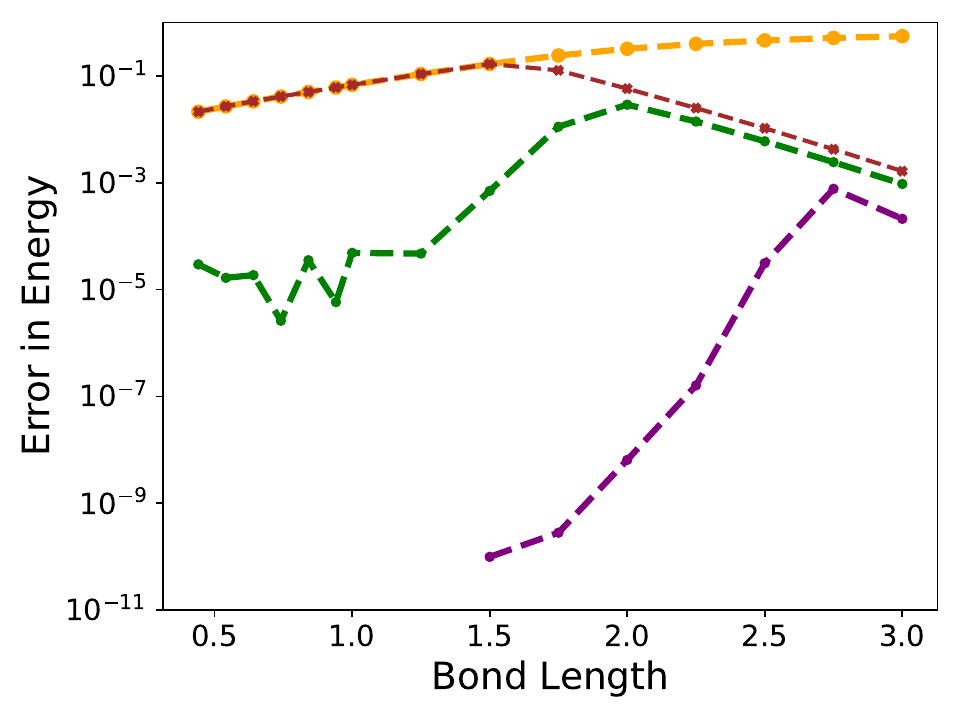} \\
    \midrule
    \textbf{b) BeH$_2$ molecule } \\
    \midrule
    \includegraphics[width=0.35\textwidth]{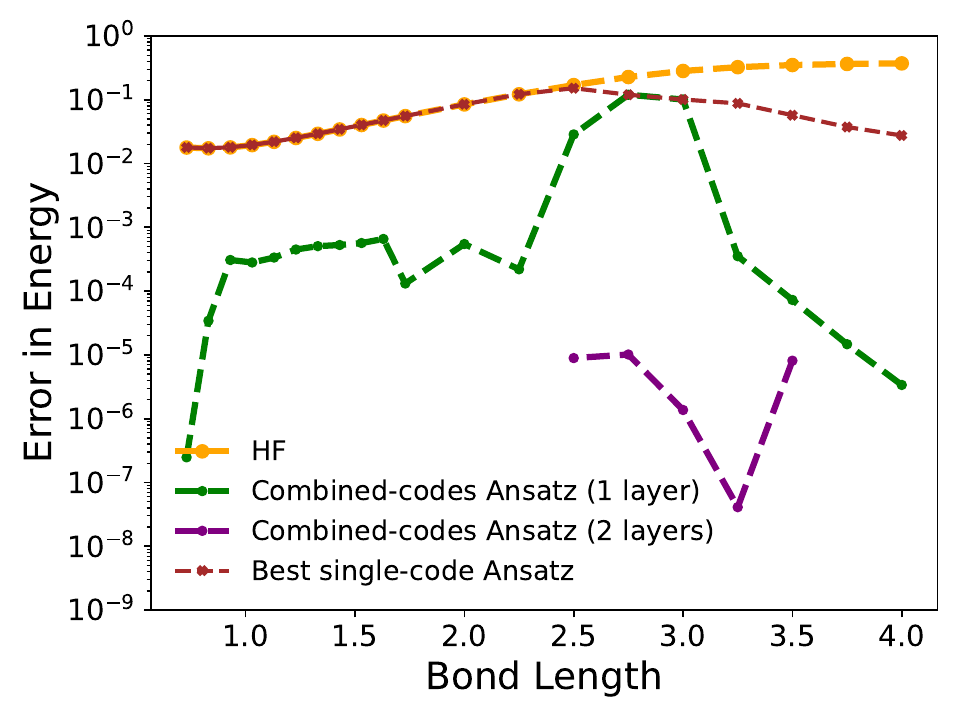} \\
    \midrule
    \textbf{c) H$_2$O molecule }\\
    \midrule
    \includegraphics[width=0.35\textwidth]{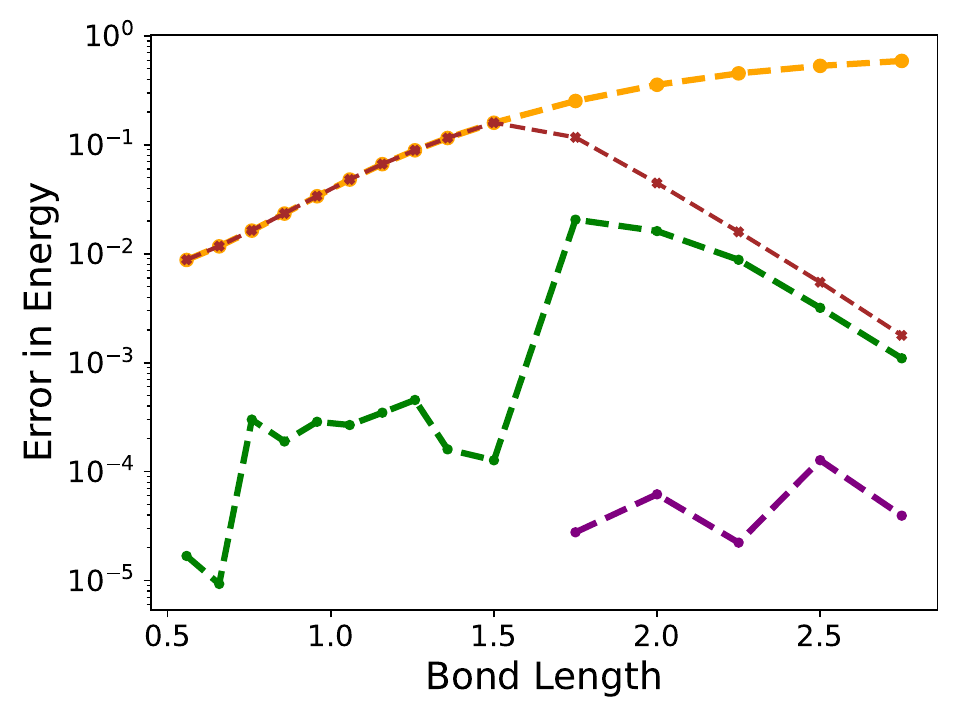} \\
    \end{tabular}
    \caption{\label{fig:BeH2} VQE results from optimization of ground state energies for H$_{4}$, BeH$_{2}$ and H$_2$O molecule using different geometries with single and two layers of the proposed ansatz.} 
\end{figure}

In this section, we perform additional simulations using two layers of the combined-codes ansatz for H$_4$, BeH$_2$ and H$_2$O molecule. 
Specifically, we focus on geometries where a single layer of the combined-codes ansatz fails to converge to the ground-state energies. 
We plot the energy errors for all the molecules obtained from these simulations in Fig.~\ref{fig:BeH2}.
For comparison, we also show the errors obtained using a single layer.

As shown in Fig.~\ref{fig:BeH2}, the results with two layers of the ansatz converge to the true ground-state energy within chemical accuracy ($10^{-3}$Ha) for geometries where a single layer fails to achieve this threshold. 
We also performed additional simulations for the N\textsubscript{2} molecule at bond distances of 1.5\AA{} and 1.75\AA{}, where the two-layer ansatz converged to lower energy errors.
At 1.5\AA{}, the error decreased from 7.054mHa to 0.99mHa, and at 1.75\AA{}, from 59.2mHa to 4.47mHa, demonstrating approximately an order of magnitude improvement in both cases.

\begin{figure}[htbp!]
    \centering
    \includegraphics[width=0.3\textwidth]{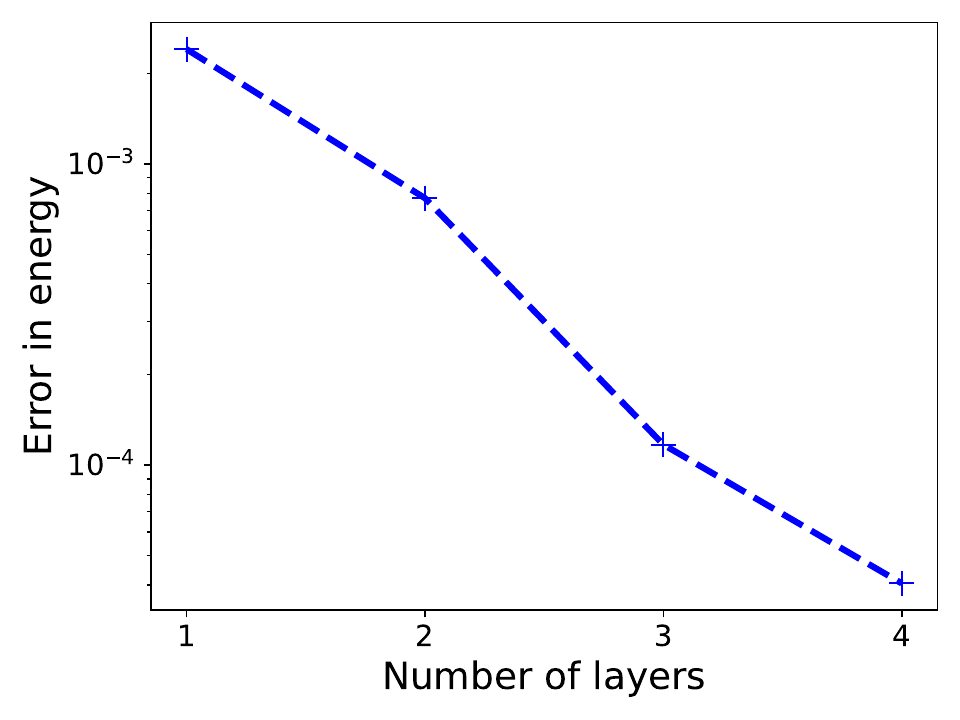} \\
    \caption{\label{fig:H4_laye} Error in the final ground-state energy for the H\textsubscript{4} molecule as a function of the number of layers in the combined codes ansatz.} 
\end{figure}

Furthermore, in Fig.~\ref{fig:H4_laye}, we plot the error in the final ground-state energy for the H\textsubscript{4} molecule at a bond length of 2.75\AA{}, as a function of the number of layers in the combined codes ansatz. 
We observe that the error systematically decreases with increasing number of layers.
However, the error here remains at $4\times 10^{-5}$Ha even with four layers, which is significantly higher compared to the errors achieved with just two layers for other geometries, as shown in Fig.~\ref{fig:BeH2}(a).
We attribute this saturation to the  limited number of optimization iterations, and expect that allowing more iterations or employing a modified optimization procedure~\cite{skolik2021layerwise, brnovic2024efficient, anand2024information} would enable further reductions in the energy error.

From all the numerical experiments discussed thus far, we conclude that the proposed ansatz effectively approximates eigenvalues of molecules of varying complexity. 
While a single layer of the ansatz performs well for small molecules, increasing the number of layers enhances its accuracy for larger molecules.
To further validate the proposed method, we compare its performance against the traditional Variational Hamiltonian Ansatz (VHA) in the following section.

\begin{figure*}[htbp!]
    \centering
    \begin{tabular}{c c c}
    \toprule
    \textbf{a) 0.79 } & \textbf{b) 1.09} & \textbf{c) 1.29} \\
    \midrule
    \includegraphics[width=0.3\textwidth]{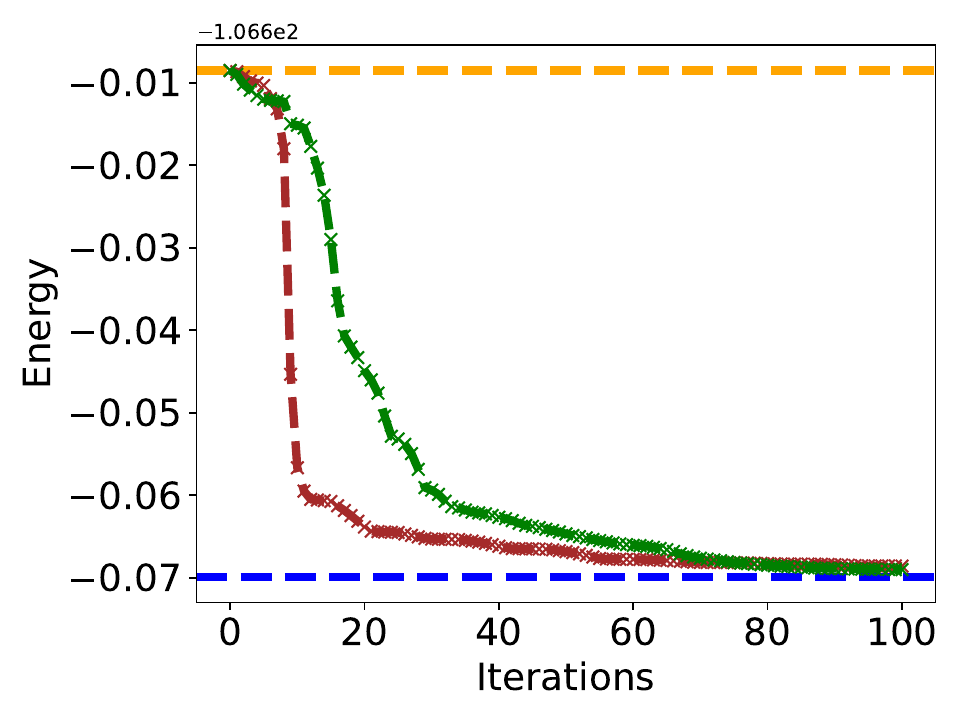} & 
    \includegraphics[width=0.3\textwidth]{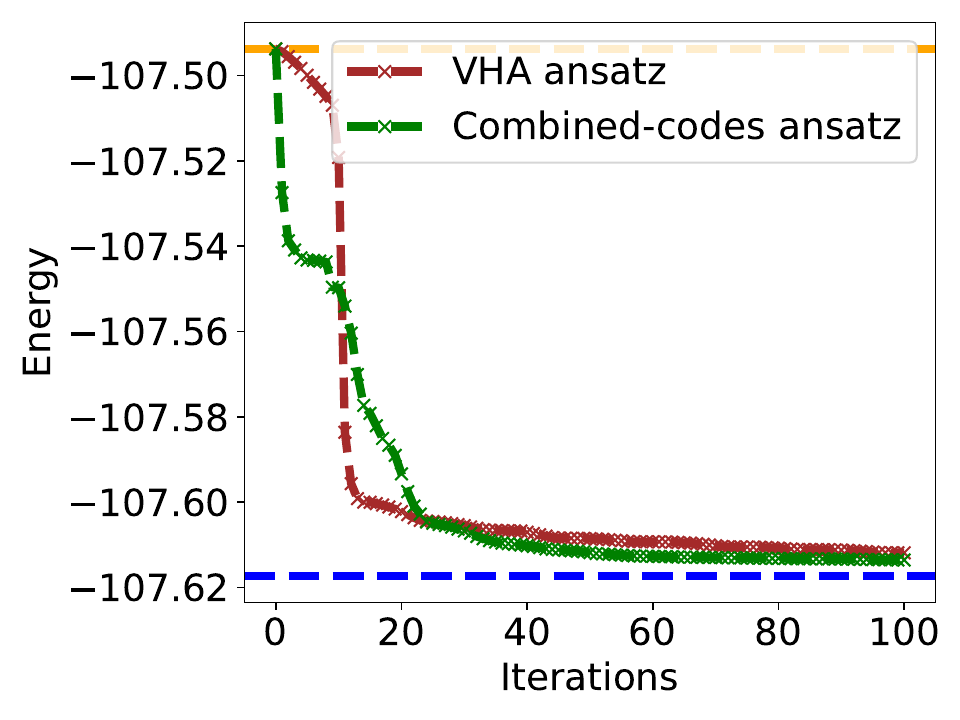} & 
    \includegraphics[width=0.3\textwidth]{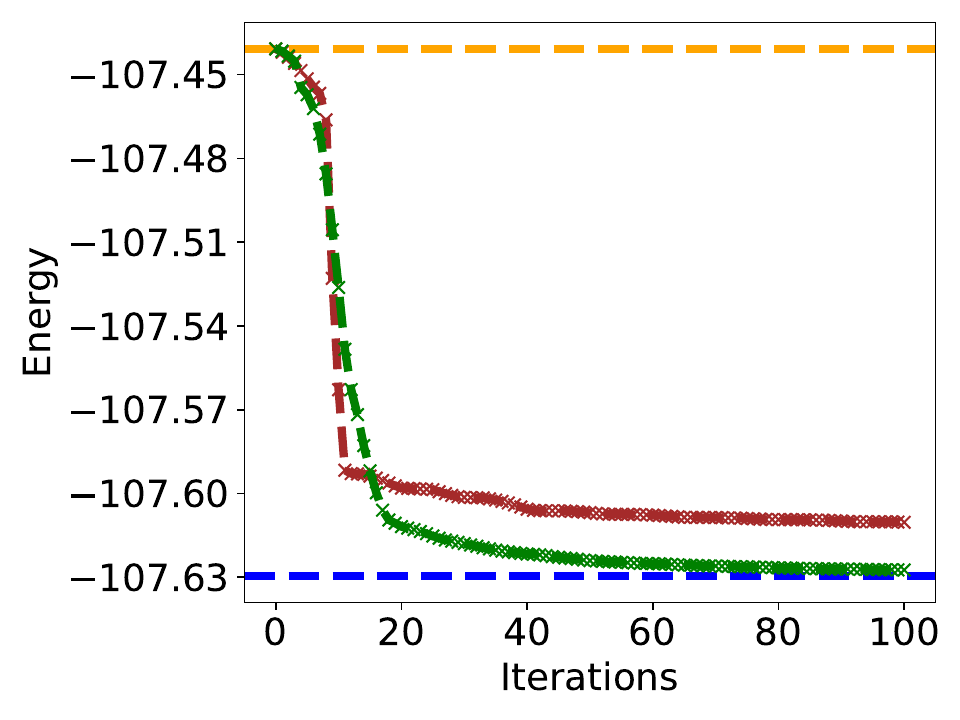} \\
    
    \end{tabular}
    \begin{tabular}{c c}
    \toprule
    \textbf{d) 1.50}  & \textbf{e) 2.0} \\
    \midrule
    \includegraphics[width=0.3\textwidth]{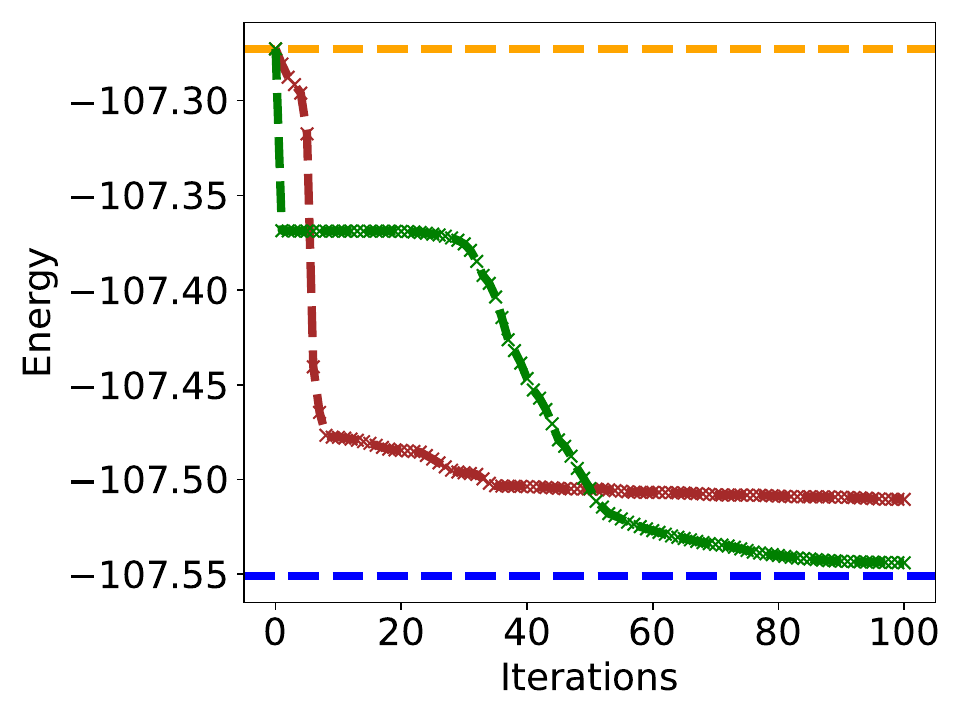} & 
    \includegraphics[width=0.3\textwidth]{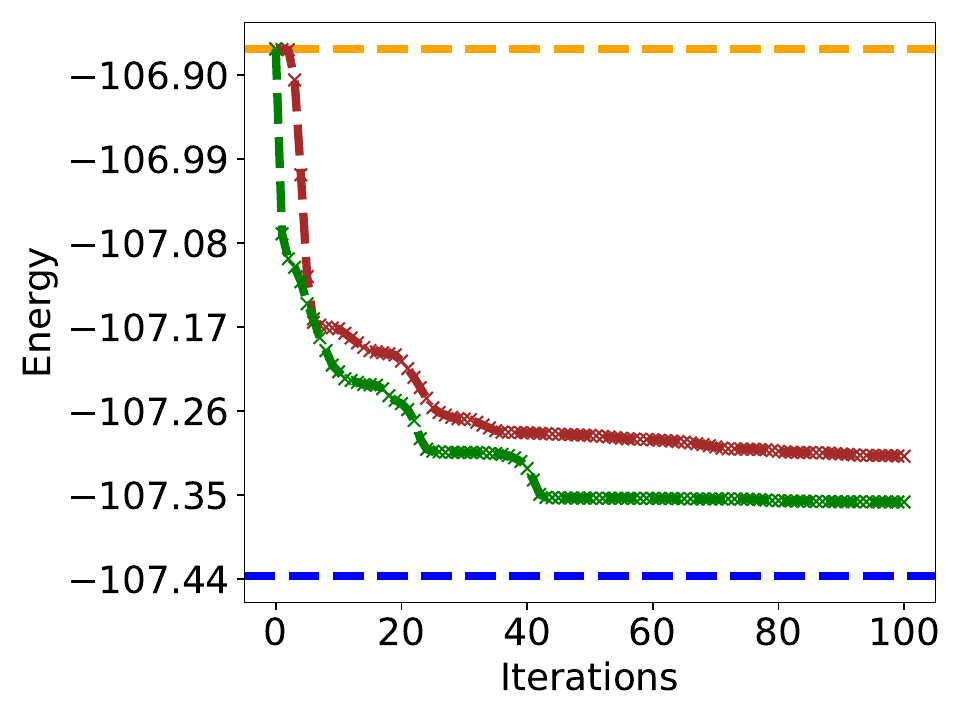} \\
  
    \end{tabular}
    \caption{\label{fig:vha_comparison} Optimization trajectories from simulations of the N$_{2}$ molecule at five distinct bond lengths (as indicated in the sub-captions) using the traditional VHA ansatz and the proposed ansatz. The orange and blue lines corresponds to Hartree-Fock and the exact ground state energies, respectively.} 
\end{figure*}

\subsection{Comparison with VHA}\label{subsec:comp_VHA}
To benchmark the performance of the proposed ansatz against the traditional VHA, we simulate the active space of the nitrogen molecule (N$_2$), as described in Section~\ref{subsubsec:H2OandN2}.
Simulations are performed for five different geometries: squeezed, equilibrium, and stretched. The results are summarized in Fig.~\ref{fig:vha_comparison}.

From Fig.~\ref{fig:vha_comparison}, we observe that the proposed ansatz converges to slightly better energy values compared to the VHA across all configurations. 
Additionally, the energy gap between the two methods increases as the geometry stretches, highlighting the superior convergence properties of the proposed ansatz for highly correlated systems.

To complement the energy optimization results, we compare the circuit complexity of the proposed and variational Hamiltonian ansatz (VHA) for all molecules studied in this work.  
Table~\ref{tab:VHAvsMVHA} provides a detailed comparison of the two-qubit gate counts and the number of parameters for both approaches. 
We present results for both a standard implementation of the VHA ansatz and an optimized implementation similar to that described in Refs.~\cite{van2020circuit, anand2025hamiltonian}.
We compile the ans\"atze into CNOT (two-qubit) and single-qubit gates and observe that the two-qubit gate counts for the combined-codes ansatz (CCA) are comparable to those of the standard VHA implementation and in some cases, even modestly lower for larger molecules. 
Although the gate counts for the CCA are similar to the standard implementation of VHA, they are significantly higher than those of the optimized implementation. 
However, we note that applying similar circuit optimizations to the CCA, as discussed in Ref.~\cite{anand2025hamiltonian}, can reduce its gate counts further, bringing them below those of the optimized VHA. 

\begin{table}[htbp!]
    \centering
    \begin{tabular}{c| c c | c c}
        \hline
        \multicolumn{1}{c}{Molecule} & \multicolumn{2}{c}{Number of 2-qubit gates} & \multicolumn{2}{c}{Number of Parameters}\\
        \hline
        \hline
        & VHA & CCA & VHA & CCA \\
        \hline
        H$_2$ & 36 (21) & 48 & 14 & 12 \\ 
        \hline
        LiH & 262 (134) & 272 & 61 & 108 \\ 
        \hline
        H$_4$ & 1328 (448) & 1254 & 184 & 197 \\ 
        \hline
        BeH$_2$ & 1328 (445) & 1272 & 184 & 192 \\ 
        \hline
        H$_2$O & 2042 (655) & 1408 & 251 & 474 \\ 
        \hline
        N$_2$ & 1860 (660) & 1740 & 246 & 362 \\ 
        \hline
        \hline
    \end{tabular}
    \caption{A table containing the gate complexity of a single layer of the VHA ansatz and the proposed combined-codes ansatz (CCA). 
    For the VHA, the optimized CNOT counts obtained using the method in Ref.~\cite{van2020circuit} are shown in parentheses.
    The values listed here are the average of all the circuits used for simulation corresponding to the different molecules.}
    \label{tab:VHAvsMVHA}
\end{table}

We also observe that the proposed ansatz requires more parameters than the traditional VHA. 
Additionally, we note that the number of parameters in the proposed ansatz is slightly higher than the number of FCI parameters, which scales as $m \choose n$, where $m$ is the number of spatial orbitals and $n$ is the number of electrons. 

Furthermore, the two-qubit gate counts for the molecules considered here are comparable to those reported for other fixed-circuit ansätze in the literature~\cite{kottmann2022optimized}, but they tend to be higher than those achieved with adaptive approaches~\cite{anastasiou2024tetris}. 
However, further modifications to the proposed circuits~\cite{anand2025hamiltonian} could reduce the gate counts, making them comparable to those of adaptive methods.

\section{Conclusion}\label{sec:conclusion}
In this work, we propose a new approach for designing quantum circuits to approximate the low-lying eigenvalues of molecular Hamiltonians. 
Our method leverages efficient clustering techniques to identify groups of mutually commuting terms in the Hamiltonian and employs Clifford unitaries to simultaneously diagonalize the operators within each cluster.
We introduce two distinct ans\"atze: the ``single-code" and ``combined-codes" ansatz, constructed using the stabilizer states associated with each set of commuting groups.

The single-code ansatz, which builds on the Hartree–Fock state, consistently outperforms the widely used Hartree–Fock state in terms of energy for all molecules considered in this study. 
Notably, a single layer of the ansatz is classically simulatable.

This result highlights its potential as a better choice for initial state preparation, particularly for molecular geometries where Hartree-Fock states are suboptimal.
These findings align with and contribute to the growing body of research~\cite{ravi2022cafqa, schleich2023partitioning, anand2022quantum} exploring practical applications with Clifford circuits.

Additionally, we present empirical evidence demonstrating the effectiveness of a single layer of the combined-codes ansatz in approximating ground-state energies of various molecular systems. 
Additionally, we show that increasing the number of layers in the ansatz significantly enhances accuracy, particularly for systems with high correlations.

Finally, we conduct a comparative analysis of the combined-codes ansatz and the traditional Variational Hamiltonian Ansatz (VHA).
Our results show that the combined-codes ansatz not only achieves slightly better convergence but also exhibits a modest reduction in gate complexity for larger molecular systems.
This advantage underscores its scalability and practical utility for quantum simulations of molecular systems.

Our research marks an initial step towards the development of quantum circuits that incorporate a combination of problem-dependent and random single qubit unitaries, enabling symmetry-breaking and potentially leading to improved convergence for various problems of interest. 
We anticipate that the presented ansatz will unlock new possibilities for exploring the applicability of the proposed ans\"atze in other areas of physics and machine learning, while also providing a solid foundation for further investigations into such types of ans\"atze.
Future investigations will focus on further reducing gate complexity~\cite{anand2025hamiltonian} and optimizing mappings to specific hardware architectures~\cite{miller2022hardware}, ensuring broader applicability and performance on real quantum devices.

The framework can also be integrated with adaptive methods~\cite{grimsley2019adaptive, ryabinkin2018qubit}, where instead of adding individual operators we adaptively add different single-code ans\"atze to construct even more compact circuits~\cite{anastasiou2024tetris}.
Additionally, the circuit design introduced here naturally supports error detection schemes~\cite{anand2024stabilizer, delfosse2024low}, which, together with recent proposals such as Ref.~\cite{khitrin2025unbiasedobservableestimationnoisy}, can further enhance its utility for near-term quantum devices.

\section*{Acknowledgements}
This work was supported by the National Science Foundation (NSF) Quantum Leap Challenge Institute of Robust Quantum Simulation (QLCI grant OMA-2120757).
A.A. also acknowledges support by the National Science Foundation (Grant No. DMS-1925919), as part of the work was done when he was visiting the Institute for Pure and Applied Mathematics (IPAM).

\bibliography{main.bib}

\begin{thebibliography}{60}%
\makeatletter
\providecommand \@ifxundefined [1]{%
 \@ifx{#1\undefined}
}%
\providecommand \@ifnum [1]{%
 \ifnum #1\expandafter \@firstoftwo
 \else \expandafter \@secondoftwo
 \fi
}%
\providecommand \@ifx [1]{%
 \ifx #1\expandafter \@firstoftwo
 \else \expandafter \@secondoftwo
 \fi
}%
\providecommand \natexlab [1]{#1}%
\providecommand \enquote  [1]{``#1''}%
\providecommand \bibnamefont  [1]{#1}%
\providecommand \bibfnamefont [1]{#1}%
\providecommand \citenamefont [1]{#1}%
\providecommand \href@noop [0]{\@secondoftwo}%
\providecommand \href [0]{\begingroup \@sanitize@url \@href}%
\providecommand \@href[1]{\@@startlink{#1}\@@href}%
\providecommand \@@href[1]{\endgroup#1\@@endlink}%
\providecommand \@sanitize@url [0]{\catcode `\\12\catcode `\$12\catcode
  `\&12\catcode `\#12\catcode `\^12\catcode `\_12\catcode `\%12\relax}%
\providecommand \@@startlink[1]{}%
\providecommand \@@endlink[0]{}%
\providecommand \url  [0]{\begingroup\@sanitize@url \@url }%
\providecommand \@url [1]{\endgroup\@href {#1}{\urlprefix }}%
\providecommand \urlprefix  [0]{URL }%
\providecommand \Eprint [0]{\href }%
\providecommand \doibase [0]{https://doi.org/}%
\providecommand \selectlanguage [0]{\@gobble}%
\providecommand \bibinfo  [0]{\@secondoftwo}%
\providecommand \bibfield  [0]{\@secondoftwo}%
\providecommand \translation [1]{[#1]}%
\providecommand \BibitemOpen [0]{}%
\providecommand \bibitemStop [0]{}%
\providecommand \bibitemNoStop [0]{.\EOS\space}%
\providecommand \EOS [0]{\spacefactor3000\relax}%
\providecommand \BibitemShut  [1]{\csname bibitem#1\endcsname}%
\let\auto@bib@innerbib\@empty
\bibitem [{\citenamefont {Bharti}\ \emph {et~al.}(2022)\citenamefont {Bharti},
  \citenamefont {Cervera-Lierta}, \citenamefont {Kyaw}, \citenamefont {Haug},
  \citenamefont {Alperin-Lea}, \citenamefont {Anand}, \citenamefont {Degroote},
  \citenamefont {Heimonen}, \citenamefont {Kottmann}, \citenamefont {Menke}
  \emph {et~al.}}]{bharti2022noisy}%
  \BibitemOpen
  \bibfield  {author} {\bibinfo {author} {\bibfnamefont {K.}~\bibnamefont
  {Bharti}}, \bibinfo {author} {\bibfnamefont {A.}~\bibnamefont
  {Cervera-Lierta}}, \bibinfo {author} {\bibfnamefont {T.~H.}\ \bibnamefont
  {Kyaw}}, \bibinfo {author} {\bibfnamefont {T.}~\bibnamefont {Haug}}, \bibinfo
  {author} {\bibfnamefont {S.}~\bibnamefont {Alperin-Lea}}, \bibinfo {author}
  {\bibfnamefont {A.}~\bibnamefont {Anand}}, \bibinfo {author} {\bibfnamefont
  {M.}~\bibnamefont {Degroote}}, \bibinfo {author} {\bibfnamefont
  {H.}~\bibnamefont {Heimonen}}, \bibinfo {author} {\bibfnamefont {J.~S.}\
  \bibnamefont {Kottmann}}, \bibinfo {author} {\bibfnamefont {T.}~\bibnamefont
  {Menke}}, \emph {et~al.},\ }\href@noop {} {\bibfield  {journal} {\bibinfo
  {journal} {Reviews of Modern Physics}\ }\textbf {\bibinfo {volume} {94}},\
  \bibinfo {pages} {015004} (\bibinfo {year} {2022})}\BibitemShut {NoStop}%
\bibitem [{\citenamefont {Cerezo}\ \emph
  {et~al.}(2021{\natexlab{a}})\citenamefont {Cerezo}, \citenamefont
  {Arrasmith}, \citenamefont {Babbush}, \citenamefont {Benjamin}, \citenamefont
  {Endo}, \citenamefont {Fujii}, \citenamefont {McClean}, \citenamefont
  {Mitarai}, \citenamefont {Yuan}, \citenamefont {Cincio} \emph
  {et~al.}}]{cerezo2020variational}%
  \BibitemOpen
  \bibfield  {author} {\bibinfo {author} {\bibfnamefont {M.}~\bibnamefont
  {Cerezo}}, \bibinfo {author} {\bibfnamefont {A.}~\bibnamefont {Arrasmith}},
  \bibinfo {author} {\bibfnamefont {R.}~\bibnamefont {Babbush}}, \bibinfo
  {author} {\bibfnamefont {S.~C.}\ \bibnamefont {Benjamin}}, \bibinfo {author}
  {\bibfnamefont {S.}~\bibnamefont {Endo}}, \bibinfo {author} {\bibfnamefont
  {K.}~\bibnamefont {Fujii}}, \bibinfo {author} {\bibfnamefont {J.~R.}\
  \bibnamefont {McClean}}, \bibinfo {author} {\bibfnamefont {K.}~\bibnamefont
  {Mitarai}}, \bibinfo {author} {\bibfnamefont {X.}~\bibnamefont {Yuan}},
  \bibinfo {author} {\bibfnamefont {L.}~\bibnamefont {Cincio}}, \emph
  {et~al.},\ }\href@noop {} {\bibfield  {journal} {\bibinfo  {journal} {Nature
  Reviews Physics}\ ,\ \bibinfo {pages} {1}} (\bibinfo {year}
  {2021}{\natexlab{a}})}\BibitemShut {NoStop}%
\bibitem [{\citenamefont {Anand}\ \emph
  {et~al.}(2022{\natexlab{a}})\citenamefont {Anand}, \citenamefont {Schleich},
  \citenamefont {Alperin-Lea}, \citenamefont {Jensen}, \citenamefont {Sim},
  \citenamefont {D{\'{i}}az-Tinoco}, \citenamefont {Kottmann}, \citenamefont
  {Degroote}, \citenamefont {Izmaylov},\ and\ \citenamefont
  {Aspuru-Guzik}}]{Anand2021Quantum}%
  \BibitemOpen
  \bibfield  {author} {\bibinfo {author} {\bibfnamefont {A.}~\bibnamefont
  {Anand}}, \bibinfo {author} {\bibfnamefont {P.}~\bibnamefont {Schleich}},
  \bibinfo {author} {\bibfnamefont {S.}~\bibnamefont {Alperin-Lea}}, \bibinfo
  {author} {\bibfnamefont {P.~W.~K.}\ \bibnamefont {Jensen}}, \bibinfo {author}
  {\bibfnamefont {S.}~\bibnamefont {Sim}}, \bibinfo {author} {\bibfnamefont
  {M.}~\bibnamefont {D{\'{i}}az-Tinoco}}, \bibinfo {author} {\bibfnamefont
  {J.~S.}\ \bibnamefont {Kottmann}}, \bibinfo {author} {\bibfnamefont
  {M.}~\bibnamefont {Degroote}}, \bibinfo {author} {\bibfnamefont {A.~F.}\
  \bibnamefont {Izmaylov}},\ and\ \bibinfo {author} {\bibfnamefont
  {A.}~\bibnamefont {Aspuru-Guzik}},\ }\bibfield  {journal} {\bibinfo
  {journal} {Chemical Society Reviews}\ }\href
  {https://doi.org/10.1039/D1CS00932J} {10.1039/D1CS00932J} (\bibinfo {year}
  {2022}{\natexlab{a}}),\ \Eprint {https://arxiv.org/abs/2109.15176}
  {arXiv:2109.15176} \BibitemShut {NoStop}%
\bibitem [{\citenamefont {McArdle}\ \emph {et~al.}(2020)\citenamefont
  {McArdle}, \citenamefont {Endo}, \citenamefont {Aspuru-Guzik}, \citenamefont
  {Benjamin},\ and\ \citenamefont {Yuan}}]{mcardle2020quantumreview}%
  \BibitemOpen
  \bibfield  {author} {\bibinfo {author} {\bibfnamefont {S.}~\bibnamefont
  {McArdle}}, \bibinfo {author} {\bibfnamefont {S.}~\bibnamefont {Endo}},
  \bibinfo {author} {\bibfnamefont {A.}~\bibnamefont {Aspuru-Guzik}}, \bibinfo
  {author} {\bibfnamefont {S.~C.}\ \bibnamefont {Benjamin}},\ and\ \bibinfo
  {author} {\bibfnamefont {X.}~\bibnamefont {Yuan}},\ }\href@noop {} {\bibfield
   {journal} {\bibinfo  {journal} {Reviews of Modern Physics}\ }\textbf
  {\bibinfo {volume} {92}},\ \bibinfo {pages} {015003} (\bibinfo {year}
  {2020})}\BibitemShut {NoStop}%
\bibitem [{\citenamefont {Feynman}(1982)}]{feynman1982simulating}%
  \BibitemOpen
  \bibfield  {author} {\bibinfo {author} {\bibfnamefont {R.~P.}\ \bibnamefont
  {Feynman}},\ }\href@noop {} {\bibfield  {journal} {\bibinfo  {journal} {Int.
  J. Theor. Phys}\ }\textbf {\bibinfo {volume} {21}} (\bibinfo {year}
  {1982})}\BibitemShut {NoStop}%
\bibitem [{\citenamefont {Aspuru-Guzik}\ \emph {et~al.}(2005)\citenamefont
  {Aspuru-Guzik}, \citenamefont {Dutoi}, \citenamefont {Love},\ and\
  \citenamefont {Head-Gordon}}]{aspuru2005simulated}%
  \BibitemOpen
  \bibfield  {author} {\bibinfo {author} {\bibfnamefont {A.}~\bibnamefont
  {Aspuru-Guzik}}, \bibinfo {author} {\bibfnamefont {A.~D.}\ \bibnamefont
  {Dutoi}}, \bibinfo {author} {\bibfnamefont {P.~J.}\ \bibnamefont {Love}},\
  and\ \bibinfo {author} {\bibfnamefont {M.}~\bibnamefont {Head-Gordon}},\
  }\href@noop {} {\bibfield  {journal} {\bibinfo  {journal} {Science}\ }\textbf
  {\bibinfo {volume} {309}},\ \bibinfo {pages} {1704} (\bibinfo {year}
  {2005})}\BibitemShut {NoStop}%
\bibitem [{\citenamefont {Preskill}(2018)}]{preskill2018quantum}%
  \BibitemOpen
  \bibfield  {author} {\bibinfo {author} {\bibfnamefont {J.}~\bibnamefont
  {Preskill}},\ }\href@noop {} {\bibfield  {journal} {\bibinfo  {journal}
  {Quantum}\ }\textbf {\bibinfo {volume} {2}},\ \bibinfo {pages} {79} (\bibinfo
  {year} {2018})}\BibitemShut {NoStop}%
\bibitem [{\citenamefont {Peruzzo}\ \emph {et~al.}(2014)\citenamefont
  {Peruzzo}, \citenamefont {McClean}, \citenamefont {Shadbolt}, \citenamefont
  {Yung}, \citenamefont {Zhou}, \citenamefont {Love}, \citenamefont
  {Aspuru-Guzik},\ and\ \citenamefont {O’brien}}]{peruzzo2014variational}%
  \BibitemOpen
  \bibfield  {author} {\bibinfo {author} {\bibfnamefont {A.}~\bibnamefont
  {Peruzzo}}, \bibinfo {author} {\bibfnamefont {J.}~\bibnamefont {McClean}},
  \bibinfo {author} {\bibfnamefont {P.}~\bibnamefont {Shadbolt}}, \bibinfo
  {author} {\bibfnamefont {M.-H.}\ \bibnamefont {Yung}}, \bibinfo {author}
  {\bibfnamefont {X.-Q.}\ \bibnamefont {Zhou}}, \bibinfo {author}
  {\bibfnamefont {P.~J.}\ \bibnamefont {Love}}, \bibinfo {author}
  {\bibfnamefont {A.}~\bibnamefont {Aspuru-Guzik}},\ and\ \bibinfo {author}
  {\bibfnamefont {J.~L.}\ \bibnamefont {O’brien}},\ }\href@noop {} {\bibfield
   {journal} {\bibinfo  {journal} {Nature communications}\ }\textbf {\bibinfo
  {volume} {5}},\ \bibinfo {pages} {4213} (\bibinfo {year} {2014})}\BibitemShut
  {NoStop}%
\bibitem [{\citenamefont {Farhi}\ \emph {et~al.}(2014)\citenamefont {Farhi},
  \citenamefont {Goldstone},\ and\ \citenamefont {Gutmann}}]{farhi2014quantum}%
  \BibitemOpen
  \bibfield  {author} {\bibinfo {author} {\bibfnamefont {E.}~\bibnamefont
  {Farhi}}, \bibinfo {author} {\bibfnamefont {J.}~\bibnamefont {Goldstone}},\
  and\ \bibinfo {author} {\bibfnamefont {S.}~\bibnamefont {Gutmann}},\
  }\href@noop {} {\bibfield  {journal} {\bibinfo  {journal} {arXiv preprint
  arXiv:1411.4028}\ } (\bibinfo {year} {2014})}\BibitemShut {NoStop}%
\bibitem [{\citenamefont {McClean}\ \emph {et~al.}(2016)\citenamefont
  {McClean}, \citenamefont {Romero}, \citenamefont {Babbush},\ and\
  \citenamefont {Aspuru-Guzik}}]{McClean2016theoryofvqe}%
  \BibitemOpen
  \bibfield  {author} {\bibinfo {author} {\bibfnamefont {J.~R.}\ \bibnamefont
  {McClean}}, \bibinfo {author} {\bibfnamefont {J.}~\bibnamefont {Romero}},
  \bibinfo {author} {\bibfnamefont {R.}~\bibnamefont {Babbush}},\ and\ \bibinfo
  {author} {\bibfnamefont {A.}~\bibnamefont {Aspuru-Guzik}},\ }\href
  {https://doi.org/10.1088/1367-2630/18/2/023023} {\bibfield  {journal}
  {\bibinfo  {journal} {New Journal of Physics}\ }\textbf {\bibinfo {volume}
  {18}},\ \bibinfo {pages} {023023} (\bibinfo {year} {2016})}\BibitemShut
  {NoStop}%
\bibitem [{\citenamefont {Sim}\ \emph {et~al.}(2019)\citenamefont {Sim},
  \citenamefont {Johnson},\ and\ \citenamefont
  {Aspuru-Guzik}}]{sim2019expressibility}%
  \BibitemOpen
  \bibfield  {author} {\bibinfo {author} {\bibfnamefont {S.}~\bibnamefont
  {Sim}}, \bibinfo {author} {\bibfnamefont {P.~D.}\ \bibnamefont {Johnson}},\
  and\ \bibinfo {author} {\bibfnamefont {A.}~\bibnamefont {Aspuru-Guzik}},\
  }\href@noop {} {\bibfield  {journal} {\bibinfo  {journal} {Advanced Quantum
  Technologies}\ }\textbf {\bibinfo {volume} {2}},\ \bibinfo {pages} {1900070}
  (\bibinfo {year} {2019})}\BibitemShut {NoStop}%
\bibitem [{\citenamefont {Benedetti}\ \emph {et~al.}(2019)\citenamefont
  {Benedetti}, \citenamefont {Lloyd}, \citenamefont {Sack},\ and\ \citenamefont
  {Fiorentini}}]{benedetti2019parameterized}%
  \BibitemOpen
  \bibfield  {author} {\bibinfo {author} {\bibfnamefont {M.}~\bibnamefont
  {Benedetti}}, \bibinfo {author} {\bibfnamefont {E.}~\bibnamefont {Lloyd}},
  \bibinfo {author} {\bibfnamefont {S.}~\bibnamefont {Sack}},\ and\ \bibinfo
  {author} {\bibfnamefont {M.}~\bibnamefont {Fiorentini}},\ }\href@noop {}
  {\bibfield  {journal} {\bibinfo  {journal} {Quantum Science and Technology}\
  }\textbf {\bibinfo {volume} {4}},\ \bibinfo {pages} {043001} (\bibinfo {year}
  {2019})}\BibitemShut {NoStop}%
\bibitem [{\citenamefont {Cong}\ \emph {et~al.}(2019)\citenamefont {Cong},
  \citenamefont {Choi},\ and\ \citenamefont {Lukin}}]{cong2019quantum}%
  \BibitemOpen
  \bibfield  {author} {\bibinfo {author} {\bibfnamefont {I.}~\bibnamefont
  {Cong}}, \bibinfo {author} {\bibfnamefont {S.}~\bibnamefont {Choi}},\ and\
  \bibinfo {author} {\bibfnamefont {M.~D.}\ \bibnamefont {Lukin}},\ }\href@noop
  {} {\bibfield  {journal} {\bibinfo  {journal} {Nature Physics}\ }\textbf
  {\bibinfo {volume} {15}},\ \bibinfo {pages} {1273} (\bibinfo {year}
  {2019})}\BibitemShut {NoStop}%
\bibitem [{\citenamefont {Zhang}\ \emph {et~al.}(2022)\citenamefont {Zhang},
  \citenamefont {Hsieh}, \citenamefont {Zhang},\ and\ \citenamefont
  {Yao}}]{zhang2022differentiable}%
  \BibitemOpen
  \bibfield  {author} {\bibinfo {author} {\bibfnamefont {S.-X.}\ \bibnamefont
  {Zhang}}, \bibinfo {author} {\bibfnamefont {C.-Y.}\ \bibnamefont {Hsieh}},
  \bibinfo {author} {\bibfnamefont {S.}~\bibnamefont {Zhang}},\ and\ \bibinfo
  {author} {\bibfnamefont {H.}~\bibnamefont {Yao}},\ }\href@noop {} {\bibfield
  {journal} {\bibinfo  {journal} {Quantum Science and Technology}\ }\textbf
  {\bibinfo {volume} {7}},\ \bibinfo {pages} {045023} (\bibinfo {year}
  {2022})}\BibitemShut {NoStop}%
\bibitem [{\citenamefont {Du}\ \emph {et~al.}(2022)\citenamefont {Du},
  \citenamefont {Huang}, \citenamefont {You}, \citenamefont {Hsieh},\ and\
  \citenamefont {Tao}}]{du2022quantum}%
  \BibitemOpen
  \bibfield  {author} {\bibinfo {author} {\bibfnamefont {Y.}~\bibnamefont
  {Du}}, \bibinfo {author} {\bibfnamefont {T.}~\bibnamefont {Huang}}, \bibinfo
  {author} {\bibfnamefont {S.}~\bibnamefont {You}}, \bibinfo {author}
  {\bibfnamefont {M.-H.}\ \bibnamefont {Hsieh}},\ and\ \bibinfo {author}
  {\bibfnamefont {D.}~\bibnamefont {Tao}},\ }\href@noop {} {\bibfield
  {journal} {\bibinfo  {journal} {npj Quantum Information}\ }\textbf {\bibinfo
  {volume} {8}},\ \bibinfo {pages} {62} (\bibinfo {year} {2022})}\BibitemShut
  {NoStop}%
\bibitem [{\citenamefont {McClean}\ \emph {et~al.}(2018)\citenamefont
  {McClean}, \citenamefont {Boixo}, \citenamefont {Smelyanskiy}, \citenamefont
  {Babbush},\ and\ \citenamefont {Neven}}]{mcclean2018barren}%
  \BibitemOpen
  \bibfield  {author} {\bibinfo {author} {\bibfnamefont {J.~R.}\ \bibnamefont
  {McClean}}, \bibinfo {author} {\bibfnamefont {S.}~\bibnamefont {Boixo}},
  \bibinfo {author} {\bibfnamefont {V.~N.}\ \bibnamefont {Smelyanskiy}},
  \bibinfo {author} {\bibfnamefont {R.}~\bibnamefont {Babbush}},\ and\ \bibinfo
  {author} {\bibfnamefont {H.}~\bibnamefont {Neven}},\ }\href@noop {}
  {\bibfield  {journal} {\bibinfo  {journal} {Nature communications}\ }\textbf
  {\bibinfo {volume} {9}},\ \bibinfo {pages} {1} (\bibinfo {year}
  {2018})}\BibitemShut {NoStop}%
\bibitem [{\citenamefont {Cerezo}\ \emph
  {et~al.}(2021{\natexlab{b}})\citenamefont {Cerezo}, \citenamefont {Sone},
  \citenamefont {Volkoff}, \citenamefont {Cincio},\ and\ \citenamefont
  {Coles}}]{cerezo2021cost}%
  \BibitemOpen
  \bibfield  {author} {\bibinfo {author} {\bibfnamefont {M.}~\bibnamefont
  {Cerezo}}, \bibinfo {author} {\bibfnamefont {A.}~\bibnamefont {Sone}},
  \bibinfo {author} {\bibfnamefont {T.}~\bibnamefont {Volkoff}}, \bibinfo
  {author} {\bibfnamefont {L.}~\bibnamefont {Cincio}},\ and\ \bibinfo {author}
  {\bibfnamefont {P.~J.}\ \bibnamefont {Coles}},\ }\href@noop {} {\bibfield
  {journal} {\bibinfo  {journal} {Nature communications}\ }\textbf {\bibinfo
  {volume} {12}},\ \bibinfo {pages} {1} (\bibinfo {year}
  {2021}{\natexlab{b}})}\BibitemShut {NoStop}%
\bibitem [{\citenamefont {Marrero}\ \emph {et~al.}(2021)\citenamefont
  {Marrero}, \citenamefont {Kieferov{\'a}},\ and\ \citenamefont
  {Wiebe}}]{marrero2021entanglement}%
  \BibitemOpen
  \bibfield  {author} {\bibinfo {author} {\bibfnamefont {C.~O.}\ \bibnamefont
  {Marrero}}, \bibinfo {author} {\bibfnamefont {M.}~\bibnamefont
  {Kieferov{\'a}}},\ and\ \bibinfo {author} {\bibfnamefont {N.}~\bibnamefont
  {Wiebe}},\ }\href@noop {} {\bibfield  {journal} {\bibinfo  {journal} {PRX
  Quantum}\ }\textbf {\bibinfo {volume} {2}},\ \bibinfo {pages} {040316}
  (\bibinfo {year} {2021})}\BibitemShut {NoStop}%
\bibitem [{\citenamefont {Wang}\ \emph {et~al.}(2021)\citenamefont {Wang},
  \citenamefont {Fontana}, \citenamefont {Cerezo}, \citenamefont {Sharma},
  \citenamefont {Sone}, \citenamefont {Cincio},\ and\ \citenamefont
  {Coles}}]{wang2021noise}%
  \BibitemOpen
  \bibfield  {author} {\bibinfo {author} {\bibfnamefont {S.}~\bibnamefont
  {Wang}}, \bibinfo {author} {\bibfnamefont {E.}~\bibnamefont {Fontana}},
  \bibinfo {author} {\bibfnamefont {M.}~\bibnamefont {Cerezo}}, \bibinfo
  {author} {\bibfnamefont {K.}~\bibnamefont {Sharma}}, \bibinfo {author}
  {\bibfnamefont {A.}~\bibnamefont {Sone}}, \bibinfo {author} {\bibfnamefont
  {L.}~\bibnamefont {Cincio}},\ and\ \bibinfo {author} {\bibfnamefont {P.~J.}\
  \bibnamefont {Coles}},\ }\href@noop {} {\bibfield  {journal} {\bibinfo
  {journal} {Nature communications}\ }\textbf {\bibinfo {volume} {12}},\
  \bibinfo {pages} {1} (\bibinfo {year} {2021})}\BibitemShut {NoStop}%
\bibitem [{\citenamefont {Haug}\ and\ \citenamefont
  {Kim}(2021)}]{haug2021optimal}%
  \BibitemOpen
  \bibfield  {author} {\bibinfo {author} {\bibfnamefont {T.}~\bibnamefont
  {Haug}}\ and\ \bibinfo {author} {\bibfnamefont {M.}~\bibnamefont {Kim}},\
  }\href@noop {} {\bibfield  {journal} {\bibinfo  {journal} {arXiv preprint
  arXiv:2104.14543}\ } (\bibinfo {year} {2021})}\BibitemShut {NoStop}%
\bibitem [{\citenamefont {Weber}\ \emph {et~al.}(2022)\citenamefont {Weber},
  \citenamefont {Anand}, \citenamefont {Cervera-Lierta}, \citenamefont
  {Kottmann}, \citenamefont {Kyaw}, \citenamefont {Li}, \citenamefont
  {Aspuru-Guzik}, \citenamefont {Zhang},\ and\ \citenamefont
  {Zhao}}]{weber2022toward}%
  \BibitemOpen
  \bibfield  {author} {\bibinfo {author} {\bibfnamefont {M.}~\bibnamefont
  {Weber}}, \bibinfo {author} {\bibfnamefont {A.}~\bibnamefont {Anand}},
  \bibinfo {author} {\bibfnamefont {A.}~\bibnamefont {Cervera-Lierta}},
  \bibinfo {author} {\bibfnamefont {J.~S.}\ \bibnamefont {Kottmann}}, \bibinfo
  {author} {\bibfnamefont {T.~H.}\ \bibnamefont {Kyaw}}, \bibinfo {author}
  {\bibfnamefont {B.}~\bibnamefont {Li}}, \bibinfo {author} {\bibfnamefont
  {A.}~\bibnamefont {Aspuru-Guzik}}, \bibinfo {author} {\bibfnamefont
  {C.}~\bibnamefont {Zhang}},\ and\ \bibinfo {author} {\bibfnamefont
  {Z.}~\bibnamefont {Zhao}},\ }\href@noop {} {\bibfield  {journal} {\bibinfo
  {journal} {Physical Review Research}\ }\textbf {\bibinfo {volume} {4}},\
  \bibinfo {pages} {033217} (\bibinfo {year} {2022})}\BibitemShut {NoStop}%
\bibitem [{\citenamefont {Wecker}\ \emph {et~al.}(2015)\citenamefont {Wecker},
  \citenamefont {Hastings},\ and\ \citenamefont {Troyer}}]{wecker2015progress}%
  \BibitemOpen
  \bibfield  {author} {\bibinfo {author} {\bibfnamefont {D.}~\bibnamefont
  {Wecker}}, \bibinfo {author} {\bibfnamefont {M.~B.}\ \bibnamefont
  {Hastings}},\ and\ \bibinfo {author} {\bibfnamefont {M.}~\bibnamefont
  {Troyer}},\ }\href@noop {} {\bibfield  {journal} {\bibinfo  {journal}
  {Physical Review A}\ }\textbf {\bibinfo {volume} {92}},\ \bibinfo {pages}
  {042303} (\bibinfo {year} {2015})}\BibitemShut {NoStop}%
\bibitem [{\citenamefont {Wiersema}\ \emph {et~al.}(2020)\citenamefont
  {Wiersema}, \citenamefont {Zhou}, \citenamefont {de~Sereville}, \citenamefont
  {Carrasquilla}, \citenamefont {Kim},\ and\ \citenamefont
  {Yuen}}]{wiersema2020exploring}%
  \BibitemOpen
  \bibfield  {author} {\bibinfo {author} {\bibfnamefont {R.}~\bibnamefont
  {Wiersema}}, \bibinfo {author} {\bibfnamefont {C.}~\bibnamefont {Zhou}},
  \bibinfo {author} {\bibfnamefont {Y.}~\bibnamefont {de~Sereville}}, \bibinfo
  {author} {\bibfnamefont {J.~F.}\ \bibnamefont {Carrasquilla}}, \bibinfo
  {author} {\bibfnamefont {Y.~B.}\ \bibnamefont {Kim}},\ and\ \bibinfo {author}
  {\bibfnamefont {H.}~\bibnamefont {Yuen}},\ }\href@noop {} {\bibfield
  {journal} {\bibinfo  {journal} {PRX Quantum}\ }\textbf {\bibinfo {volume}
  {1}},\ \bibinfo {pages} {020319} (\bibinfo {year} {2020})}\BibitemShut
  {NoStop}%
\bibitem [{\citenamefont {Choquette}\ \emph {et~al.}(2021)\citenamefont
  {Choquette}, \citenamefont {Di~Paolo}, \citenamefont {Barkoutsos},
  \citenamefont {S{\'e}n{\'e}chal}, \citenamefont {Tavernelli},\ and\
  \citenamefont {Blais}}]{choquette2020quantum}%
  \BibitemOpen
  \bibfield  {author} {\bibinfo {author} {\bibfnamefont {A.}~\bibnamefont
  {Choquette}}, \bibinfo {author} {\bibfnamefont {A.}~\bibnamefont {Di~Paolo}},
  \bibinfo {author} {\bibfnamefont {P.~K.}\ \bibnamefont {Barkoutsos}},
  \bibinfo {author} {\bibfnamefont {D.}~\bibnamefont {S{\'e}n{\'e}chal}},
  \bibinfo {author} {\bibfnamefont {I.}~\bibnamefont {Tavernelli}},\ and\
  \bibinfo {author} {\bibfnamefont {A.}~\bibnamefont {Blais}},\ }\href@noop {}
  {\bibfield  {journal} {\bibinfo  {journal} {Physical Review Research}\
  }\textbf {\bibinfo {volume} {3}},\ \bibinfo {pages} {023092} (\bibinfo {year}
  {2021})}\BibitemShut {NoStop}%
\bibitem [{\citenamefont {Anand}\ \emph
  {et~al.}(2022{\natexlab{b}})\citenamefont {Anand}, \citenamefont
  {Alperin-Lea}, \citenamefont {Choquette},\ and\ \citenamefont
  {Aspuru-Guzik}}]{anand2022exploring}%
  \BibitemOpen
  \bibfield  {author} {\bibinfo {author} {\bibfnamefont {A.}~\bibnamefont
  {Anand}}, \bibinfo {author} {\bibfnamefont {S.}~\bibnamefont {Alperin-Lea}},
  \bibinfo {author} {\bibfnamefont {A.}~\bibnamefont {Choquette}},\ and\
  \bibinfo {author} {\bibfnamefont {A.}~\bibnamefont {Aspuru-Guzik}},\
  }\href@noop {} {\bibfield  {journal} {\bibinfo  {journal} {arXiv preprint
  arXiv:2209.14405}\ } (\bibinfo {year} {2022}{\natexlab{b}})}\BibitemShut
  {NoStop}%
\bibitem [{\citenamefont {Aaronson}\ and\ \citenamefont
  {Gottesman}(2004)}]{aaronson2004improved}%
  \BibitemOpen
  \bibfield  {author} {\bibinfo {author} {\bibfnamefont {S.}~\bibnamefont
  {Aaronson}}\ and\ \bibinfo {author} {\bibfnamefont {D.}~\bibnamefont
  {Gottesman}},\ }\href@noop {} {\bibfield  {journal} {\bibinfo  {journal}
  {Physical Review A}\ }\textbf {\bibinfo {volume} {70}},\ \bibinfo {pages}
  {052328} (\bibinfo {year} {2004})}\BibitemShut {NoStop}%
\bibitem [{\citenamefont {Bravyi}\ and\ \citenamefont
  {Gosset}(2016)}]{bravyi2016improved}%
  \BibitemOpen
  \bibfield  {author} {\bibinfo {author} {\bibfnamefont {S.}~\bibnamefont
  {Bravyi}}\ and\ \bibinfo {author} {\bibfnamefont {D.}~\bibnamefont
  {Gosset}},\ }\href@noop {} {\bibfield  {journal} {\bibinfo  {journal}
  {Physical review letters}\ }\textbf {\bibinfo {volume} {116}},\ \bibinfo
  {pages} {250501} (\bibinfo {year} {2016})}\BibitemShut {NoStop}%
\bibitem [{\citenamefont {Verteletskyi}\ \emph {et~al.}(2020)\citenamefont
  {Verteletskyi}, \citenamefont {Yen},\ and\ \citenamefont
  {Izmaylov}}]{verteletskyi2020measurement}%
  \BibitemOpen
  \bibfield  {author} {\bibinfo {author} {\bibfnamefont {V.}~\bibnamefont
  {Verteletskyi}}, \bibinfo {author} {\bibfnamefont {T.-C.}\ \bibnamefont
  {Yen}},\ and\ \bibinfo {author} {\bibfnamefont {A.~F.}\ \bibnamefont
  {Izmaylov}},\ }\href@noop {} {\bibfield  {journal} {\bibinfo  {journal} {The
  Journal of chemical physics}\ }\textbf {\bibinfo {volume} {152}} (\bibinfo
  {year} {2020})}\BibitemShut {NoStop}%
\bibitem [{\citenamefont {Izmaylov}\ \emph {et~al.}(2019)\citenamefont
  {Izmaylov}, \citenamefont {Yen}, \citenamefont {Lang},\ and\ \citenamefont
  {Verteletskyi}}]{izmaylov2019unitary}%
  \BibitemOpen
  \bibfield  {author} {\bibinfo {author} {\bibfnamefont {A.~F.}\ \bibnamefont
  {Izmaylov}}, \bibinfo {author} {\bibfnamefont {T.-C.}\ \bibnamefont {Yen}},
  \bibinfo {author} {\bibfnamefont {R.~A.}\ \bibnamefont {Lang}},\ and\
  \bibinfo {author} {\bibfnamefont {V.}~\bibnamefont {Verteletskyi}},\
  }\href@noop {} {\bibfield  {journal} {\bibinfo  {journal} {Journal of
  chemical theory and computation}\ }\textbf {\bibinfo {volume} {16}},\
  \bibinfo {pages} {190} (\bibinfo {year} {2019})}\BibitemShut {NoStop}%
\bibitem [{\citenamefont {Jena}\ \emph {et~al.}(2019)\citenamefont {Jena},
  \citenamefont {Genin},\ and\ \citenamefont {Mosca}}]{jena2019pauli}%
  \BibitemOpen
  \bibfield  {author} {\bibinfo {author} {\bibfnamefont {A.}~\bibnamefont
  {Jena}}, \bibinfo {author} {\bibfnamefont {S.}~\bibnamefont {Genin}},\ and\
  \bibinfo {author} {\bibfnamefont {M.}~\bibnamefont {Mosca}},\ }\href@noop {}
  {\bibfield  {journal} {\bibinfo  {journal} {arXiv preprint arXiv:1907.07859}\
  } (\bibinfo {year} {2019})}\BibitemShut {NoStop}%
\bibitem [{\citenamefont {Crawford}\ \emph {et~al.}(2021)\citenamefont
  {Crawford}, \citenamefont {van Straaten}, \citenamefont {Wang}, \citenamefont
  {Parks}, \citenamefont {Campbell},\ and\ \citenamefont
  {Brierley}}]{crawford2021efficient}%
  \BibitemOpen
  \bibfield  {author} {\bibinfo {author} {\bibfnamefont {O.}~\bibnamefont
  {Crawford}}, \bibinfo {author} {\bibfnamefont {B.}~\bibnamefont {van
  Straaten}}, \bibinfo {author} {\bibfnamefont {D.}~\bibnamefont {Wang}},
  \bibinfo {author} {\bibfnamefont {T.}~\bibnamefont {Parks}}, \bibinfo
  {author} {\bibfnamefont {E.}~\bibnamefont {Campbell}},\ and\ \bibinfo
  {author} {\bibfnamefont {S.}~\bibnamefont {Brierley}},\ }\href@noop {}
  {\bibfield  {journal} {\bibinfo  {journal} {Quantum}\ }\textbf {\bibinfo
  {volume} {5}},\ \bibinfo {pages} {385} (\bibinfo {year} {2021})}\BibitemShut
  {NoStop}%
\bibitem [{\citenamefont {Huggins}\ \emph {et~al.}(2021)\citenamefont
  {Huggins}, \citenamefont {McClean}, \citenamefont {Rubin}, \citenamefont
  {Jiang}, \citenamefont {Wiebe}, \citenamefont {Whaley},\ and\ \citenamefont
  {Babbush}}]{huggins2021efficient}%
  \BibitemOpen
  \bibfield  {author} {\bibinfo {author} {\bibfnamefont {W.~J.}\ \bibnamefont
  {Huggins}}, \bibinfo {author} {\bibfnamefont {J.~R.}\ \bibnamefont
  {McClean}}, \bibinfo {author} {\bibfnamefont {N.~C.}\ \bibnamefont {Rubin}},
  \bibinfo {author} {\bibfnamefont {Z.}~\bibnamefont {Jiang}}, \bibinfo
  {author} {\bibfnamefont {N.}~\bibnamefont {Wiebe}}, \bibinfo {author}
  {\bibfnamefont {K.~B.}\ \bibnamefont {Whaley}},\ and\ \bibinfo {author}
  {\bibfnamefont {R.}~\bibnamefont {Babbush}},\ }\href@noop {} {\bibfield
  {journal} {\bibinfo  {journal} {npj Quantum Information}\ }\textbf {\bibinfo
  {volume} {7}},\ \bibinfo {pages} {23} (\bibinfo {year} {2021})}\BibitemShut
  {NoStop}%
\bibitem [{\citenamefont {Gokhale}\ \emph {et~al.}(2020)\citenamefont
  {Gokhale}, \citenamefont {Angiuli}, \citenamefont {Ding}, \citenamefont
  {Gui}, \citenamefont {Tomesh}, \citenamefont {Suchara}, \citenamefont
  {Martonosi},\ and\ \citenamefont {Chong}}]{gokhale2020n}%
  \BibitemOpen
  \bibfield  {author} {\bibinfo {author} {\bibfnamefont {P.}~\bibnamefont
  {Gokhale}}, \bibinfo {author} {\bibfnamefont {O.}~\bibnamefont {Angiuli}},
  \bibinfo {author} {\bibfnamefont {Y.}~\bibnamefont {Ding}}, \bibinfo {author}
  {\bibfnamefont {K.}~\bibnamefont {Gui}}, \bibinfo {author} {\bibfnamefont
  {T.}~\bibnamefont {Tomesh}}, \bibinfo {author} {\bibfnamefont
  {M.}~\bibnamefont {Suchara}}, \bibinfo {author} {\bibfnamefont
  {M.}~\bibnamefont {Martonosi}},\ and\ \bibinfo {author} {\bibfnamefont
  {F.~T.}\ \bibnamefont {Chong}},\ }\href@noop {} {\bibfield  {journal}
  {\bibinfo  {journal} {IEEE Transactions on Quantum Engineering}\ }\textbf
  {\bibinfo {volume} {1}},\ \bibinfo {pages} {1} (\bibinfo {year}
  {2020})}\BibitemShut {NoStop}%
\bibitem [{\citenamefont {Zhao}\ \emph {et~al.}(2020)\citenamefont {Zhao},
  \citenamefont {Tranter}, \citenamefont {Kirby}, \citenamefont {Ung},
  \citenamefont {Miyake},\ and\ \citenamefont {Love}}]{zhao2020measurement}%
  \BibitemOpen
  \bibfield  {author} {\bibinfo {author} {\bibfnamefont {A.}~\bibnamefont
  {Zhao}}, \bibinfo {author} {\bibfnamefont {A.}~\bibnamefont {Tranter}},
  \bibinfo {author} {\bibfnamefont {W.~M.}\ \bibnamefont {Kirby}}, \bibinfo
  {author} {\bibfnamefont {S.~F.}\ \bibnamefont {Ung}}, \bibinfo {author}
  {\bibfnamefont {A.}~\bibnamefont {Miyake}},\ and\ \bibinfo {author}
  {\bibfnamefont {P.~J.}\ \bibnamefont {Love}},\ }\href@noop {} {\bibfield
  {journal} {\bibinfo  {journal} {Physical Review A}\ }\textbf {\bibinfo
  {volume} {101}},\ \bibinfo {pages} {062322} (\bibinfo {year}
  {2020})}\BibitemShut {NoStop}%
\bibitem [{\citenamefont {Anand}\ \emph
  {et~al.}(2022{\natexlab{c}})\citenamefont {Anand}, \citenamefont {Kottmann},\
  and\ \citenamefont {Aspuru-Guzik}}]{anand2022quantum}%
  \BibitemOpen
  \bibfield  {author} {\bibinfo {author} {\bibfnamefont {A.}~\bibnamefont
  {Anand}}, \bibinfo {author} {\bibfnamefont {J.~S.}\ \bibnamefont
  {Kottmann}},\ and\ \bibinfo {author} {\bibfnamefont {A.}~\bibnamefont
  {Aspuru-Guzik}},\ }\href@noop {} {\bibfield  {journal} {\bibinfo  {journal}
  {arXiv preprint arXiv:2207.02961}\ } (\bibinfo {year}
  {2022}{\natexlab{c}})}\BibitemShut {NoStop}%
\bibitem [{\citenamefont {Schleich}\ \emph {et~al.}(2023)\citenamefont
  {Schleich}, \citenamefont {Boen}, \citenamefont {Cincio}, \citenamefont
  {Anand}, \citenamefont {Kottmann}, \citenamefont {Tretiak}, \citenamefont
  {Dub},\ and\ \citenamefont {Aspuru-Guzik}}]{schleich2023partitioning}%
  \BibitemOpen
  \bibfield  {author} {\bibinfo {author} {\bibfnamefont {P.}~\bibnamefont
  {Schleich}}, \bibinfo {author} {\bibfnamefont {J.}~\bibnamefont {Boen}},
  \bibinfo {author} {\bibfnamefont {L.}~\bibnamefont {Cincio}}, \bibinfo
  {author} {\bibfnamefont {A.}~\bibnamefont {Anand}}, \bibinfo {author}
  {\bibfnamefont {J.~S.}\ \bibnamefont {Kottmann}}, \bibinfo {author}
  {\bibfnamefont {S.}~\bibnamefont {Tretiak}}, \bibinfo {author} {\bibfnamefont
  {P.~A.}\ \bibnamefont {Dub}},\ and\ \bibinfo {author} {\bibfnamefont
  {A.}~\bibnamefont {Aspuru-Guzik}},\ }\href@noop {} {\bibfield  {journal}
  {\bibinfo  {journal} {Journal of Chemical Theory and Computation}\ }\textbf
  {\bibinfo {volume} {19}},\ \bibinfo {pages} {4952} (\bibinfo {year}
  {2023})}\BibitemShut {NoStop}%
\bibitem [{\citenamefont {Ravi}\ \emph {et~al.}(2022)\citenamefont {Ravi},
  \citenamefont {Gokhale}, \citenamefont {Ding}, \citenamefont {Kirby},
  \citenamefont {Smith}, \citenamefont {Baker}, \citenamefont {Love},
  \citenamefont {Hoffmann}, \citenamefont {Brown},\ and\ \citenamefont
  {Chong}}]{ravi2022cafqa}%
  \BibitemOpen
  \bibfield  {author} {\bibinfo {author} {\bibfnamefont {G.~S.}\ \bibnamefont
  {Ravi}}, \bibinfo {author} {\bibfnamefont {P.}~\bibnamefont {Gokhale}},
  \bibinfo {author} {\bibfnamefont {Y.}~\bibnamefont {Ding}}, \bibinfo {author}
  {\bibfnamefont {W.}~\bibnamefont {Kirby}}, \bibinfo {author} {\bibfnamefont
  {K.}~\bibnamefont {Smith}}, \bibinfo {author} {\bibfnamefont {J.~M.}\
  \bibnamefont {Baker}}, \bibinfo {author} {\bibfnamefont {P.~J.}\ \bibnamefont
  {Love}}, \bibinfo {author} {\bibfnamefont {H.}~\bibnamefont {Hoffmann}},
  \bibinfo {author} {\bibfnamefont {K.~R.}\ \bibnamefont {Brown}},\ and\
  \bibinfo {author} {\bibfnamefont {F.~T.}\ \bibnamefont {Chong}},\ }in\
  \href@noop {} {\emph {\bibinfo {booktitle} {Proceedings of the 28th ACM
  International Conference on Architectural Support for Programming Languages
  and Operating Systems, Volume 1}}}\ (\bibinfo {year} {2022})\ pp.\ \bibinfo
  {pages} {15--29}\BibitemShut {NoStop}%
\bibitem [{\citenamefont {Anand}\ and\ \citenamefont
  {Brown}(2025)}]{anand2025hamiltonian}%
  \BibitemOpen
  \bibfield  {author} {\bibinfo {author} {\bibfnamefont {A.}~\bibnamefont
  {Anand}}\ and\ \bibinfo {author} {\bibfnamefont {K.~R.}\ \bibnamefont
  {Brown}},\ }\href@noop {} {\bibfield  {journal} {\bibinfo  {journal}
  {Physical Review A}\ }\textbf {\bibinfo {volume} {111}},\ \bibinfo {pages}
  {012437} (\bibinfo {year} {2025})}\BibitemShut {NoStop}%
\bibitem [{\citenamefont {Suzuki}(1976)}]{suzuki1976generalized}%
  \BibitemOpen
  \bibfield  {author} {\bibinfo {author} {\bibfnamefont {M.}~\bibnamefont
  {Suzuki}},\ }\href {https://doi.org/10.1007/BF01609348} {\bibfield  {journal}
  {\bibinfo  {journal} {Commun. Math. Phys.}\ }\textbf {\bibinfo {volume}
  {51}},\ \bibinfo {pages} {183} (\bibinfo {year} {1976})}\BibitemShut
  {NoStop}%
\bibitem [{\citenamefont {Gottesman}(1997)}]{gottesman1997stabilizer}%
  \BibitemOpen
  \bibfield  {author} {\bibinfo {author} {\bibfnamefont {D.}~\bibnamefont
  {Gottesman}},\ }\href@noop {} {\emph {\bibinfo {title} {Stabilizer codes and
  quantum error correction}}}\ (\bibinfo  {publisher} {California Institute of
  Technology},\ \bibinfo {year} {1997})\BibitemShut {NoStop}%
\bibitem [{\citenamefont {Ryabinkin}\ \emph
  {et~al.}(2018{\natexlab{a}})\citenamefont {Ryabinkin}, \citenamefont
  {Genin},\ and\ \citenamefont {Izmaylov}}]{ryabinkin2018constrained}%
  \BibitemOpen
  \bibfield  {author} {\bibinfo {author} {\bibfnamefont {I.~G.}\ \bibnamefont
  {Ryabinkin}}, \bibinfo {author} {\bibfnamefont {S.~N.}\ \bibnamefont
  {Genin}},\ and\ \bibinfo {author} {\bibfnamefont {A.~F.}\ \bibnamefont
  {Izmaylov}},\ }\href@noop {} {\bibfield  {journal} {\bibinfo  {journal}
  {Journal of chemical theory and computation}\ }\textbf {\bibinfo {volume}
  {15}},\ \bibinfo {pages} {249} (\bibinfo {year}
  {2018}{\natexlab{a}})}\BibitemShut {NoStop}%
\bibitem [{\citenamefont {Tang}\ \emph {et~al.}(2021)\citenamefont {Tang},
  \citenamefont {Shkolnikov}, \citenamefont {Barron}, \citenamefont {Grimsley},
  \citenamefont {Mayhall}, \citenamefont {Barnes},\ and\ \citenamefont
  {Economou}}]{tang2021qubit}%
  \BibitemOpen
  \bibfield  {author} {\bibinfo {author} {\bibfnamefont {H.~L.}\ \bibnamefont
  {Tang}}, \bibinfo {author} {\bibfnamefont {V.}~\bibnamefont {Shkolnikov}},
  \bibinfo {author} {\bibfnamefont {G.~S.}\ \bibnamefont {Barron}}, \bibinfo
  {author} {\bibfnamefont {H.~R.}\ \bibnamefont {Grimsley}}, \bibinfo {author}
  {\bibfnamefont {N.~J.}\ \bibnamefont {Mayhall}}, \bibinfo {author}
  {\bibfnamefont {E.}~\bibnamefont {Barnes}},\ and\ \bibinfo {author}
  {\bibfnamefont {S.~E.}\ \bibnamefont {Economou}},\ }\href@noop {} {\bibfield
  {journal} {\bibinfo  {journal} {PRX Quantum}\ }\textbf {\bibinfo {volume}
  {2}},\ \bibinfo {pages} {020310} (\bibinfo {year} {2021})}\BibitemShut
  {NoStop}%
\bibitem [{\citenamefont {Grimsley}\ \emph {et~al.}(2019)\citenamefont
  {Grimsley}, \citenamefont {Economou}, \citenamefont {Barnes},\ and\
  \citenamefont {Mayhall}}]{grimsley2019adaptive}%
  \BibitemOpen
  \bibfield  {author} {\bibinfo {author} {\bibfnamefont {H.~R.}\ \bibnamefont
  {Grimsley}}, \bibinfo {author} {\bibfnamefont {S.~E.}\ \bibnamefont
  {Economou}}, \bibinfo {author} {\bibfnamefont {E.}~\bibnamefont {Barnes}},\
  and\ \bibinfo {author} {\bibfnamefont {N.~J.}\ \bibnamefont {Mayhall}},\
  }\href@noop {} {\bibfield  {journal} {\bibinfo  {journal} {Nature
  communications}\ }\textbf {\bibinfo {volume} {10}},\ \bibinfo {pages} {3007}
  (\bibinfo {year} {2019})}\BibitemShut {NoStop}%
\bibitem [{\citenamefont {Ryabinkin}\ \emph
  {et~al.}(2018{\natexlab{b}})\citenamefont {Ryabinkin}, \citenamefont {Yen},
  \citenamefont {Genin},\ and\ \citenamefont {Izmaylov}}]{ryabinkin2018qubit}%
  \BibitemOpen
  \bibfield  {author} {\bibinfo {author} {\bibfnamefont {I.~G.}\ \bibnamefont
  {Ryabinkin}}, \bibinfo {author} {\bibfnamefont {T.-C.}\ \bibnamefont {Yen}},
  \bibinfo {author} {\bibfnamefont {S.~N.}\ \bibnamefont {Genin}},\ and\
  \bibinfo {author} {\bibfnamefont {A.~F.}\ \bibnamefont {Izmaylov}},\
  }\href@noop {} {\bibfield  {journal} {\bibinfo  {journal} {Journal of
  chemical theory and computation}\ }\textbf {\bibinfo {volume} {14}},\
  \bibinfo {pages} {6317} (\bibinfo {year} {2018}{\natexlab{b}})}\BibitemShut
  {NoStop}%
\bibitem [{\citenamefont {Ryabinkin}\ \emph {et~al.}(2020)\citenamefont
  {Ryabinkin}, \citenamefont {Lang}, \citenamefont {Genin},\ and\ \citenamefont
  {Izmaylov}}]{ryabinkin2020iterative}%
  \BibitemOpen
  \bibfield  {author} {\bibinfo {author} {\bibfnamefont {I.~G.}\ \bibnamefont
  {Ryabinkin}}, \bibinfo {author} {\bibfnamefont {R.~A.}\ \bibnamefont {Lang}},
  \bibinfo {author} {\bibfnamefont {S.~N.}\ \bibnamefont {Genin}},\ and\
  \bibinfo {author} {\bibfnamefont {A.~F.}\ \bibnamefont {Izmaylov}},\
  }\href@noop {} {\bibfield  {journal} {\bibinfo  {journal} {Journal of
  chemical theory and computation}\ }\textbf {\bibinfo {volume} {16}},\
  \bibinfo {pages} {1055} (\bibinfo {year} {2020})}\BibitemShut {NoStop}%
\bibitem [{\citenamefont {Van Den~Berg}\ and\ \citenamefont
  {Temme}(2020)}]{van2020circuit}%
  \BibitemOpen
  \bibfield  {author} {\bibinfo {author} {\bibfnamefont {E.}~\bibnamefont {Van
  Den~Berg}}\ and\ \bibinfo {author} {\bibfnamefont {K.}~\bibnamefont
  {Temme}},\ }\href@noop {} {\bibfield  {journal} {\bibinfo  {journal}
  {Quantum}\ }\textbf {\bibinfo {volume} {4}},\ \bibinfo {pages} {322}
  (\bibinfo {year} {2020})}\BibitemShut {NoStop}%
\bibitem [{\citenamefont {Kottmann}\ \emph {et~al.}(2021)\citenamefont
  {Kottmann}, \citenamefont {Alperin-Lea}, \citenamefont {Tamayo-Mendoza},
  \citenamefont {Cervera-Lierta}, \citenamefont {Lavigne}, \citenamefont {Yen},
  \citenamefont {Verteletskyi}, \citenamefont {Schleich}, \citenamefont
  {Anand}, \citenamefont {Degroote} \emph {et~al.}}]{kottmann2021tequila}%
  \BibitemOpen
  \bibfield  {author} {\bibinfo {author} {\bibfnamefont {J.~S.}\ \bibnamefont
  {Kottmann}}, \bibinfo {author} {\bibfnamefont {S.}~\bibnamefont
  {Alperin-Lea}}, \bibinfo {author} {\bibfnamefont {T.}~\bibnamefont
  {Tamayo-Mendoza}}, \bibinfo {author} {\bibfnamefont {A.}~\bibnamefont
  {Cervera-Lierta}}, \bibinfo {author} {\bibfnamefont {C.}~\bibnamefont
  {Lavigne}}, \bibinfo {author} {\bibfnamefont {T.-C.}\ \bibnamefont {Yen}},
  \bibinfo {author} {\bibfnamefont {V.}~\bibnamefont {Verteletskyi}}, \bibinfo
  {author} {\bibfnamefont {P.}~\bibnamefont {Schleich}}, \bibinfo {author}
  {\bibfnamefont {A.}~\bibnamefont {Anand}}, \bibinfo {author} {\bibfnamefont
  {M.}~\bibnamefont {Degroote}}, \emph {et~al.},\ }\href@noop {} {\bibfield
  {journal} {\bibinfo  {journal} {Quantum Science and Technology}\ }\textbf
  {\bibinfo {volume} {6}},\ \bibinfo {pages} {024009} (\bibinfo {year}
  {2021})}\BibitemShut {NoStop}%
\bibitem [{\citenamefont {Suzuki}\ \emph {et~al.}(2021)\citenamefont {Suzuki},
  \citenamefont {Kawase}, \citenamefont {Masumura}, \citenamefont {Hiraga},
  \citenamefont {Nakadai}, \citenamefont {Chen}, \citenamefont {Nakanishi},
  \citenamefont {Mitarai}, \citenamefont {Imai}, \citenamefont {Tamiya} \emph
  {et~al.}}]{suzuki2021qulacs}%
  \BibitemOpen
  \bibfield  {author} {\bibinfo {author} {\bibfnamefont {Y.}~\bibnamefont
  {Suzuki}}, \bibinfo {author} {\bibfnamefont {Y.}~\bibnamefont {Kawase}},
  \bibinfo {author} {\bibfnamefont {Y.}~\bibnamefont {Masumura}}, \bibinfo
  {author} {\bibfnamefont {Y.}~\bibnamefont {Hiraga}}, \bibinfo {author}
  {\bibfnamefont {M.}~\bibnamefont {Nakadai}}, \bibinfo {author} {\bibfnamefont
  {J.}~\bibnamefont {Chen}}, \bibinfo {author} {\bibfnamefont {K.~M.}\
  \bibnamefont {Nakanishi}}, \bibinfo {author} {\bibfnamefont {K.}~\bibnamefont
  {Mitarai}}, \bibinfo {author} {\bibfnamefont {R.}~\bibnamefont {Imai}},
  \bibinfo {author} {\bibfnamefont {S.}~\bibnamefont {Tamiya}}, \emph
  {et~al.},\ }\href@noop {} {\bibfield  {journal} {\bibinfo  {journal}
  {Quantum}\ }\textbf {\bibinfo {volume} {5}},\ \bibinfo {pages} {559}
  (\bibinfo {year} {2021})}\BibitemShut {NoStop}%
\bibitem [{\citenamefont {Virtanen}\ \emph {et~al.}(2020)\citenamefont
  {Virtanen}, \citenamefont {Gommers}, \citenamefont {Oliphant}, \citenamefont
  {Haberland}, \citenamefont {Reddy}, \citenamefont {Cournapeau}, \citenamefont
  {Burovski}, \citenamefont {Peterson}, \citenamefont {Weckesser},
  \citenamefont {Bright} \emph {et~al.}}]{virtanen2020scipy}%
  \BibitemOpen
  \bibfield  {author} {\bibinfo {author} {\bibfnamefont {P.}~\bibnamefont
  {Virtanen}}, \bibinfo {author} {\bibfnamefont {R.}~\bibnamefont {Gommers}},
  \bibinfo {author} {\bibfnamefont {T.~E.}\ \bibnamefont {Oliphant}}, \bibinfo
  {author} {\bibfnamefont {M.}~\bibnamefont {Haberland}}, \bibinfo {author}
  {\bibfnamefont {T.}~\bibnamefont {Reddy}}, \bibinfo {author} {\bibfnamefont
  {D.}~\bibnamefont {Cournapeau}}, \bibinfo {author} {\bibfnamefont
  {E.}~\bibnamefont {Burovski}}, \bibinfo {author} {\bibfnamefont
  {P.}~\bibnamefont {Peterson}}, \bibinfo {author} {\bibfnamefont
  {W.}~\bibnamefont {Weckesser}}, \bibinfo {author} {\bibfnamefont
  {J.}~\bibnamefont {Bright}}, \emph {et~al.},\ }\href@noop {} {\bibfield
  {journal} {\bibinfo  {journal} {Nature methods}\ }\textbf {\bibinfo {volume}
  {17}},\ \bibinfo {pages} {261} (\bibinfo {year} {2020})}\BibitemShut
  {NoStop}%
\bibitem [{\citenamefont {Jordan}\ and\ \citenamefont
  {Wigner}(1928)}]{jordan1928pauli}%
  \BibitemOpen
  \bibfield  {author} {\bibinfo {author} {\bibfnamefont {P.}~\bibnamefont
  {Jordan}}\ and\ \bibinfo {author} {\bibfnamefont {E.~P.}\ \bibnamefont
  {Wigner}},\ }\href {https://doi.org/10.1007/BF01331938} {\bibfield  {journal}
  {\bibinfo  {journal} {Z. Phys.}\ }\textbf {\bibinfo {volume} {47}},\ \bibinfo
  {pages} {14} (\bibinfo {year} {1928})}\BibitemShut {NoStop}%
\bibitem [{git()}]{github_Ham_MVHA}%
  \BibitemOpen
  \href@noop {} {}\bibinfo {howpublished}
  {\url{https://github.com/AbhinavUofT/Modified_VHA}}\BibitemShut {NoStop}%
\bibitem [{\citenamefont {Skolik}\ \emph {et~al.}(2021)\citenamefont {Skolik},
  \citenamefont {McClean}, \citenamefont {Mohseni}, \citenamefont {Van
  Der~Smagt},\ and\ \citenamefont {Leib}}]{skolik2021layerwise}%
  \BibitemOpen
  \bibfield  {author} {\bibinfo {author} {\bibfnamefont {A.}~\bibnamefont
  {Skolik}}, \bibinfo {author} {\bibfnamefont {J.~R.}\ \bibnamefont {McClean}},
  \bibinfo {author} {\bibfnamefont {M.}~\bibnamefont {Mohseni}}, \bibinfo
  {author} {\bibfnamefont {P.}~\bibnamefont {Van Der~Smagt}},\ and\ \bibinfo
  {author} {\bibfnamefont {M.}~\bibnamefont {Leib}},\ }\href@noop {} {\bibfield
   {journal} {\bibinfo  {journal} {Quantum Machine Intelligence}\ }\textbf
  {\bibinfo {volume} {3}},\ \bibinfo {pages} {5} (\bibinfo {year}
  {2021})}\BibitemShut {NoStop}%
\bibitem [{\citenamefont {Brnovi{\'c}}\ \emph {et~al.}(2024)\citenamefont
  {Brnovi{\'c}}, \citenamefont {Iouchtchenko},\ and\ \citenamefont
  {Koch-Janusz}}]{brnovic2024efficient}%
  \BibitemOpen
  \bibfield  {author} {\bibinfo {author} {\bibfnamefont {M.}~\bibnamefont
  {Brnovi{\'c}}}, \bibinfo {author} {\bibfnamefont {D.}~\bibnamefont
  {Iouchtchenko}},\ and\ \bibinfo {author} {\bibfnamefont {M.}~\bibnamefont
  {Koch-Janusz}},\ }in\ \href@noop {} {\emph {\bibinfo {booktitle} {2024 IEEE
  International Conference on Quantum Computing and Engineering (QCE)}}},\
  Vol.~\bibinfo {volume} {2}\ (\bibinfo {organization} {IEEE},\ \bibinfo {year}
  {2024})\ pp.\ \bibinfo {pages} {571--572}\BibitemShut {NoStop}%
\bibitem [{\citenamefont {Anand}\ \emph {et~al.}(2024)\citenamefont {Anand},
  \citenamefont {Kristensen}, \citenamefont {Frohnert}, \citenamefont {Sim},\
  and\ \citenamefont {Aspuru-Guzik}}]{anand2024information}%
  \BibitemOpen
  \bibfield  {author} {\bibinfo {author} {\bibfnamefont {A.}~\bibnamefont
  {Anand}}, \bibinfo {author} {\bibfnamefont {L.~B.}\ \bibnamefont
  {Kristensen}}, \bibinfo {author} {\bibfnamefont {F.}~\bibnamefont
  {Frohnert}}, \bibinfo {author} {\bibfnamefont {S.}~\bibnamefont {Sim}},\ and\
  \bibinfo {author} {\bibfnamefont {A.}~\bibnamefont {Aspuru-Guzik}},\
  }\href@noop {} {\bibfield  {journal} {\bibinfo  {journal} {Quantum Science
  and Technology}\ }\textbf {\bibinfo {volume} {9}},\ \bibinfo {pages} {035025}
  (\bibinfo {year} {2024})}\BibitemShut {NoStop}%
\bibitem [{\citenamefont {Kottmann}\ and\ \citenamefont
  {Aspuru-Guzik}(2022)}]{kottmann2022optimized}%
  \BibitemOpen
  \bibfield  {author} {\bibinfo {author} {\bibfnamefont {J.~S.}\ \bibnamefont
  {Kottmann}}\ and\ \bibinfo {author} {\bibfnamefont {A.}~\bibnamefont
  {Aspuru-Guzik}},\ }\href@noop {} {\bibfield  {journal} {\bibinfo  {journal}
  {Physical Review A}\ }\textbf {\bibinfo {volume} {105}},\ \bibinfo {pages}
  {032449} (\bibinfo {year} {2022})}\BibitemShut {NoStop}%
\bibitem [{\citenamefont {Anastasiou}\ \emph {et~al.}(2024)\citenamefont
  {Anastasiou}, \citenamefont {Chen}, \citenamefont {Mayhall}, \citenamefont
  {Barnes},\ and\ \citenamefont {Economou}}]{anastasiou2024tetris}%
  \BibitemOpen
  \bibfield  {author} {\bibinfo {author} {\bibfnamefont {P.~G.}\ \bibnamefont
  {Anastasiou}}, \bibinfo {author} {\bibfnamefont {Y.}~\bibnamefont {Chen}},
  \bibinfo {author} {\bibfnamefont {N.~J.}\ \bibnamefont {Mayhall}}, \bibinfo
  {author} {\bibfnamefont {E.}~\bibnamefont {Barnes}},\ and\ \bibinfo {author}
  {\bibfnamefont {S.~E.}\ \bibnamefont {Economou}},\ }\href@noop {} {\bibfield
  {journal} {\bibinfo  {journal} {Physical Review Research}\ }\textbf {\bibinfo
  {volume} {6}},\ \bibinfo {pages} {013254} (\bibinfo {year}
  {2024})}\BibitemShut {NoStop}%
\bibitem [{\citenamefont {Miller}\ \emph {et~al.}(2022)\citenamefont {Miller},
  \citenamefont {Fischer}, \citenamefont {Sokolov}, \citenamefont
  {Barkoutsos},\ and\ \citenamefont {Tavernelli}}]{miller2022hardware}%
  \BibitemOpen
  \bibfield  {author} {\bibinfo {author} {\bibfnamefont {D.}~\bibnamefont
  {Miller}}, \bibinfo {author} {\bibfnamefont {L.~E.}\ \bibnamefont {Fischer}},
  \bibinfo {author} {\bibfnamefont {I.~O.}\ \bibnamefont {Sokolov}}, \bibinfo
  {author} {\bibfnamefont {P.~K.}\ \bibnamefont {Barkoutsos}},\ and\ \bibinfo
  {author} {\bibfnamefont {I.}~\bibnamefont {Tavernelli}},\ }\href@noop {}
  {\bibfield  {journal} {\bibinfo  {journal} {arXiv preprint arXiv:2203.03646}\
  } (\bibinfo {year} {2022})}\BibitemShut {NoStop}%
\bibitem [{\citenamefont {Anand}\ and\ \citenamefont
  {Brown}(2024)}]{anand2024stabilizer}%
  \BibitemOpen
  \bibfield  {author} {\bibinfo {author} {\bibfnamefont {A.}~\bibnamefont
  {Anand}}\ and\ \bibinfo {author} {\bibfnamefont {K.~R.}\ \bibnamefont
  {Brown}},\ }\href@noop {} {\bibfield  {journal} {\bibinfo  {journal} {arXiv
  preprint arXiv:2410.21125}\ } (\bibinfo {year} {2024})}\BibitemShut {NoStop}%
\bibitem [{\citenamefont {Delfosse}\ and\ \citenamefont
  {Tham}(2024)}]{delfosse2024low}%
  \BibitemOpen
  \bibfield  {author} {\bibinfo {author} {\bibfnamefont {N.}~\bibnamefont
  {Delfosse}}\ and\ \bibinfo {author} {\bibfnamefont {E.}~\bibnamefont
  {Tham}},\ }\href@noop {} {\bibfield  {journal} {\bibinfo  {journal} {arXiv
  preprint arXiv:2407.06583}\ } (\bibinfo {year} {2024})}\BibitemShut {NoStop}%
\bibitem [{\citenamefont {Khitrin}\ \emph {et~al.}(2025)\citenamefont
  {Khitrin}, \citenamefont {Brown},\ and\ \citenamefont
  {Anand}}]{khitrin2025unbiasedobservableestimationnoisy}%
  \BibitemOpen
  \bibfield  {author} {\bibinfo {author} {\bibfnamefont {D.}~\bibnamefont
  {Khitrin}}, \bibinfo {author} {\bibfnamefont {K.~R.}\ \bibnamefont {Brown}},\
  and\ \bibinfo {author} {\bibfnamefont {A.}~\bibnamefont {Anand}},\
  }\href@noop {} {\bibfield  {journal} {\bibinfo  {journal} {arXiv preprint
  arXiv:2505.11486}\ } (\bibinfo {year} {2025})}\BibitemShut {NoStop}%
\end{thebibliography}%

\clearpage
\newpage
\onecolumngrid

\section*{Appendix}\label{sec:appendix}

\subsection*{Hydrogen molecule example}
In this section we present the detailed construction of the ansatz proposed in this work using the example of hydrogen molecule in the minimal basis.
The Hamiltonian of H$_2$ in the minimal basis has 15 terms, and is:
\begin{align}
    \nonumber 
    \hat{H} &= -0.0984\hat{I}+0.1713\hat{Z}(0)+0.1713\hat{Z}(1) -0.2230\hat{Z}(2) -0.2230\hat{Z}(3)+0.1686\hat{Z}(0)\hat{Z}(1) \\ \nonumber
    & +0.1206\hat{Z}(0)\hat{Z}(2)+0.1659\hat{Z}(0)\hat{Z}(3) +0.1659\hat{Z}(1)\hat{Z}(2)+0.1206\hat{Z}(1)\hat{Z}(3)\\ \nonumber
    & +0.1744\hat{Z}(2)\hat{Z}(3) +0.0453\hat{Y}(0)\hat{X}(1)\hat{X}(2)\hat{Y}(3) -0.0453Y(0)Y(1)\hat{X}(2)\hat{X}(3)\\ \nonumber
    & -0.0453\hat{X}(0)\hat{X}(1)\hat{Y}(2)\hat{Y}(3) +0.0453\hat{X}0)\hat{Y}(1)\hat{Y}2)\hat{X}(3)
\end{align}
We can divide the Hamiltonian in two sets of commuting terms:
\begin{align}
\nonumber
    \mathcal{G}_1 &= \{ -0.0984\hat{I}, 0.1713\hat{Z}(0), 0.1713\hat{Z}(1),  -0.2230\hat{Z}(2), -0.2230\hat{Z}(3), 0.1686\hat{Z}(0)\hat{Z}(1),  0.1206\hat{Z}(0)\hat{Z}(2), \\ \nonumber & 0.1659\hat{Z}(0)\hat{Z}(3),  0.1659\hat{Z}(1)\hat{Z}(2), 0.1206\hat{Z}(1)\hat{Z}(3), 0.1744\hat{Z}(2)\hat{Z}(3) \}, \text{      and}
\end{align}

\begin{align}
\nonumber
    \mathcal{G}_2 &= \{ 0.0453\hat{Y}(0)\hat{X}(1)\hat{X}(2)\hat{Y}(3),  -0.0453Y(0)Y(1)\hat{X}(2)\hat{X}(3), -0.0453\hat{X}(0)\hat{X}(1)\hat{Y}(2)\hat{Y}(3),\\ \nonumber
    & 0.0453\hat{X}0)\hat{Y}(1)\hat{Y}2)\hat{X}(3)) \}.
\end{align}

\begin{figure}[htbp!]
    \centering
    \includegraphics[width=0.7\textwidth]{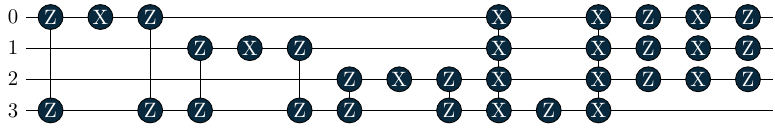}
    \caption{\label{fig:diag_h2} A Clifford unitary that diagonalizes the set $\mathcal{G}_2$. The gates are represented in a compact notation, where, a gate labeled Z(0)Z(3) == $e^{-i(\pi/2)\hat{Z}(0)\hat{Z}(3)}$ } 
\end{figure}

We can then find the Clifford unitaries that diagonalizes these groups. 
The set $\mathcal{G}_1$ is already diagonal and the unitary, $\mathcal{U}_{diag}$, that diagonalizes the set $\mathcal{G}_2$ is shown in Fig.~\ref{fig:diag_h2}.

The corresponding stabilizer groups can be written as
\begin{align}
\nonumber
    \mathcal{S}_1 &= \{ \hat{I}, -\hat{Z}(0), -\hat{Z}(1), \hat{Z}(2), \hat{Z}(3), \hat{Z}(0)\hat{Z}(1), -\hat{Z}(0)\hat{Z}(2), -\hat{Z}(0)\hat{Z}(3), -\hat{Z}(1)\hat{Z}(2), \\ \nonumber & -\hat{Z}(1)\hat{Z}(3),\hat{Z}(2)\hat{Z}(3) \}, \text{      and}
\end{align}

\begin{align}
\nonumber
    \mathcal{S}_2 &= \{ \hat{Y}(0)\hat{X}(1)\hat{X}(2)\hat{Y}(3), -Y(0)Y(1)\hat{X}(2)\hat{X}(3), -\hat{X}(0)\hat{X}(1)\hat{Y}(2)\hat{Y}(3), \hat{X}0)\hat{Y}(1)\hat{Y}2)\hat{X}(3)) \}.
\end{align}
The corresponding stabilizer states for the two groups are:
\begin{align}
    \nonumber
    \ket{\Psi_{s_1}} & = \ket{1100}\\ \nonumber
    \ket{\Psi_{s_2}} & = \mathcal{U}_{diag}^{\dagger} \ket{1100}\\ \nonumber
    & = \frac{1}{\sqrt{2}}(\ket{1100} + \ket{0011})
\end{align}

We can now use these to construct the single code ans\"atze for the two groups,

\begin{align}
\nonumber
    \ket{\Psi_1(\boldsymbol{\theta}_1)} &= U_{s{_1}}(\boldsymbol{\theta}_1)\ket{1100} \\ \nonumber
    & = \bigotimes_{j=1}^{4} \textbf{Rx}_{j}(\theta_{x_{1,j}})\textbf{Ry}_{j}(\theta_{y_{1,j}})\textbf{Rz}_{j}(\theta_{z_{1,j}}) \ket{1100}, \text{ and}
\end{align}

\begin{align}
\nonumber
    \ket{\Psi_2(\boldsymbol{\theta}_2)} &= U_{s{_2}}(\boldsymbol{\theta}_2)(\ket{1100}) \\ \nonumber
    & = \mathcal{U}_{diag}^{\dagger} \bigotimes_{j=1}^{4} \textbf{Rx}_{j}(\theta_{x_{2,j}})\textbf{Ry}_{j}(\theta_{y_{2,j}})\textbf{Rz}_{j}(\theta_{z_{2,j}}) \ket{1100}.
\end{align}

The combined codes ansatz can be then constructed as:
\begin{align}
    \nonumber
    &\ket{\Psi(\boldsymbol{\theta})}  = \mathcal{U}_{diag}^{\dagger} \bigotimes_{j=1}^{4} \textbf{Rx}_{j}(\theta_{x_{2,j}})\textbf{Ry}_{j}(\theta_{y_{2,j}})\textbf{Rz}_{j}(\theta_{z_{2,j}}) \mathcal{U}_{diag} \bigotimes_{j=1}^{4} \textbf{Rx}_{j}(\theta_{x_{1,j}}) \textbf{Ry}_{j}(\theta_{y_{1,j}}) \textbf{Rz}_{j}(\theta_{z_{1,j}}) \ket{1100}
\end{align}

\subsection*{Single-Code ansatz}
\begin{figure*}[htbp!]
    \centering
    \begin{tabular}{c c c}
    \toprule
    \textbf{a) H$_{2}$ molecule} & \textbf{b) LiH molecule} & \textbf{c) H$_{4}$ molecule} \\
    \midrule
    \includegraphics[width=0.32\textwidth]{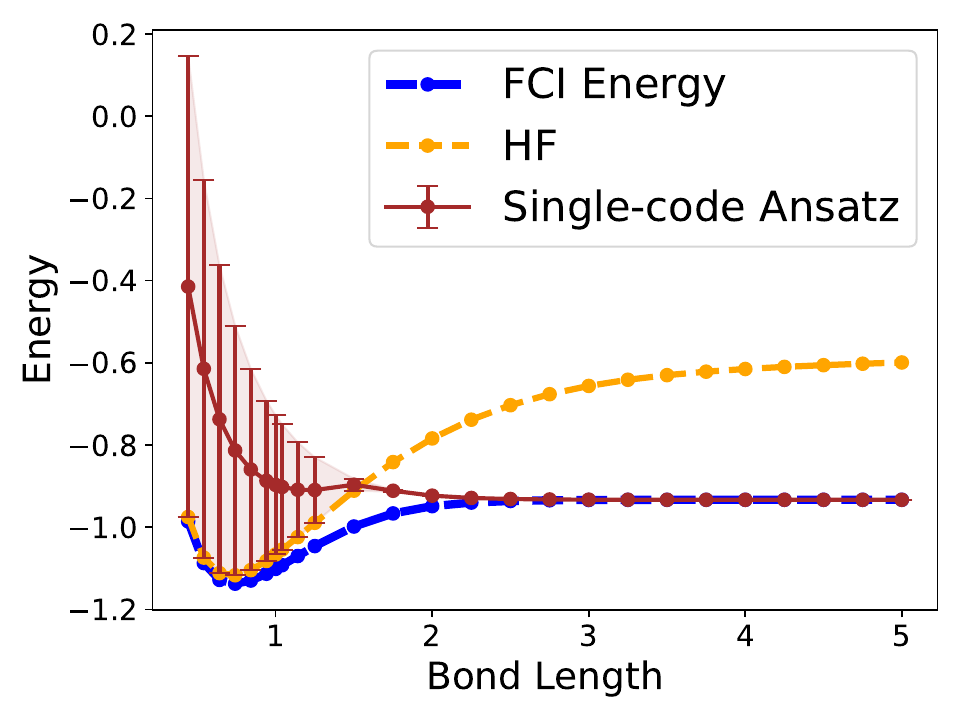} & 
    \includegraphics[width=0.32\textwidth]{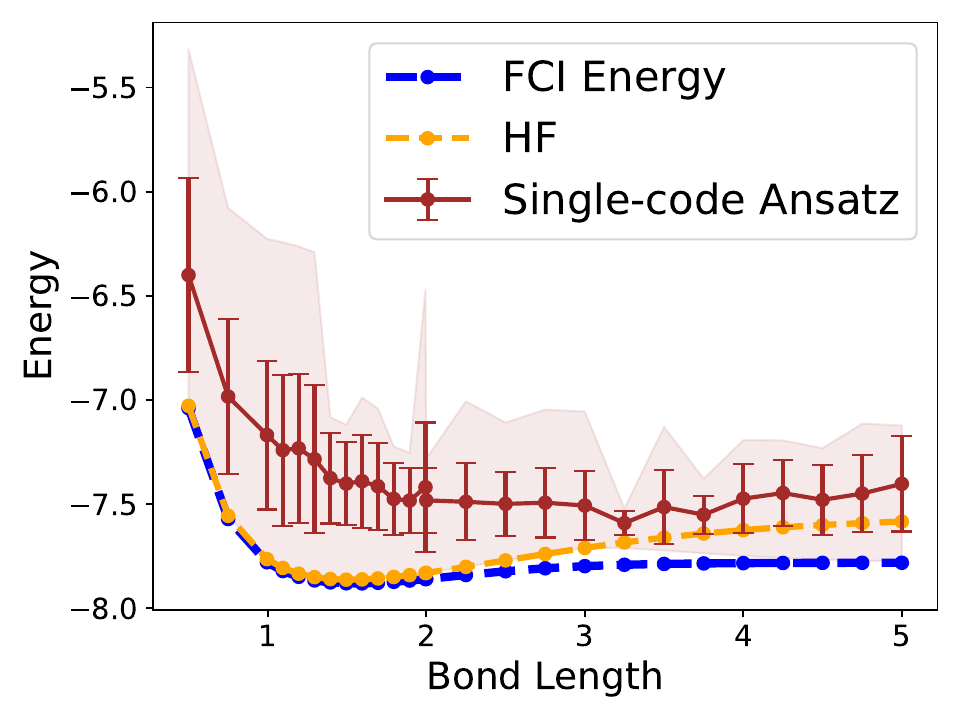} &
    \includegraphics[width=0.32\textwidth]{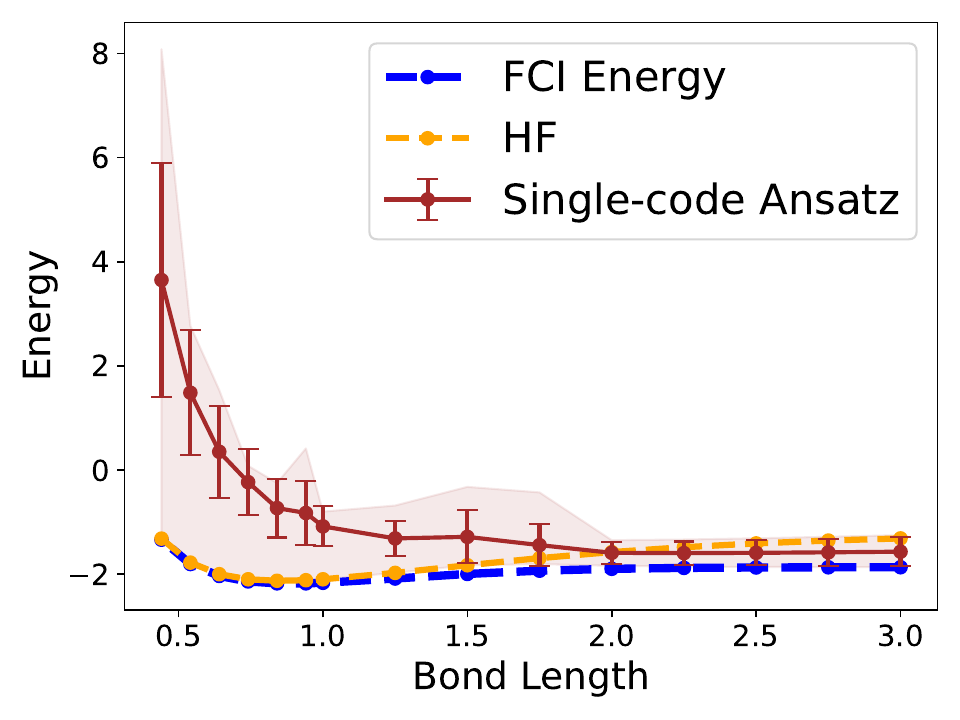}\\
    
    \end{tabular}
    \begin{tabular}{c c c}
        \midrule
        \textbf{d) BeH$_{2}$ molecule} & \textbf{e)  H$_{2}$O molecule} & \textbf{f) N$_{2}$ molecule} \\
        \midrule
        \includegraphics[width=0.32\textwidth]{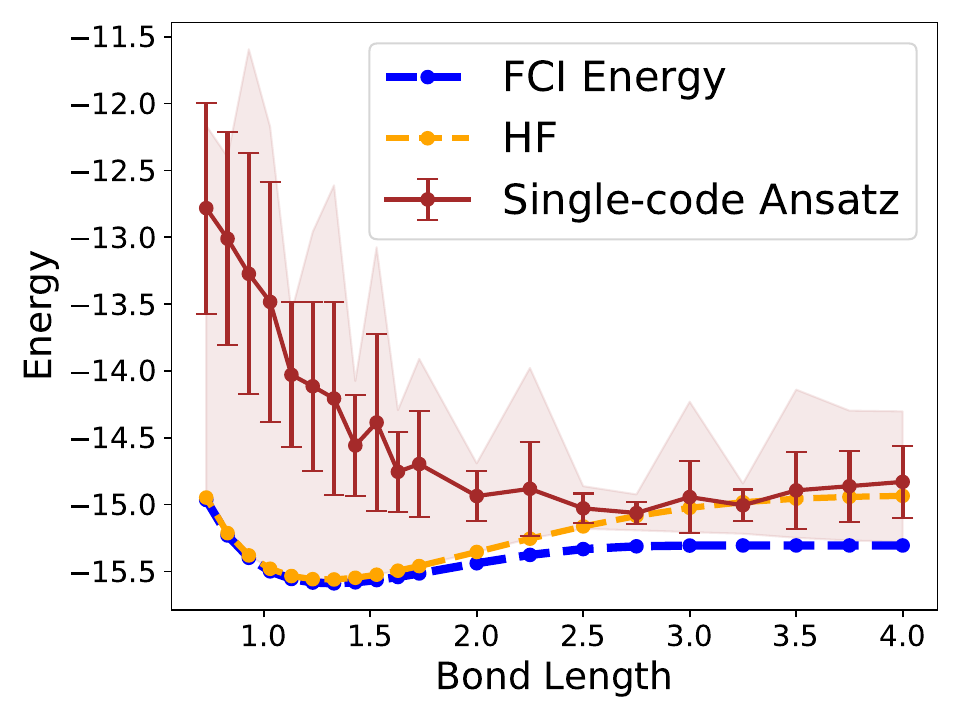} &  
         \includegraphics[width=0.32\textwidth]{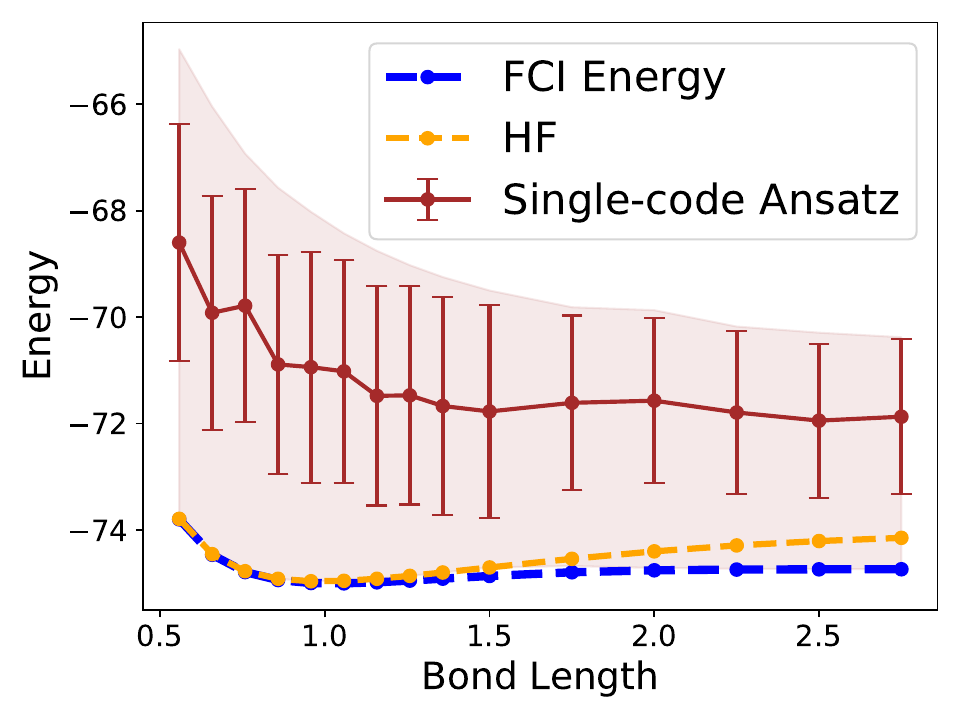} &  
         \includegraphics[width=0.32\textwidth]{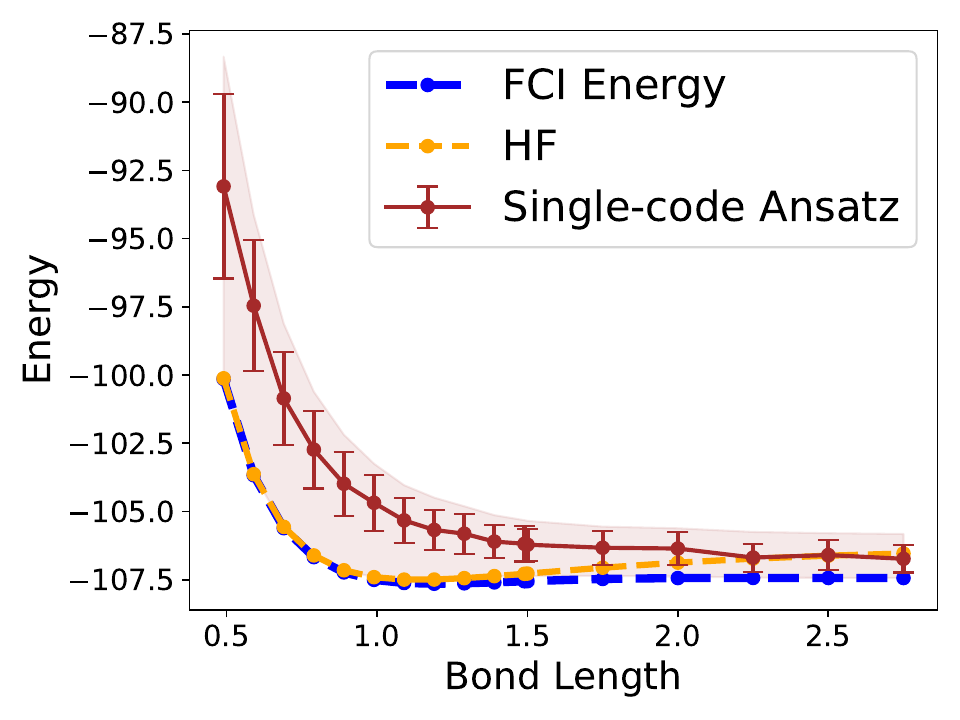} \\ 
         \midrule
    \end{tabular}
    
    \caption{\label{fig:SCAs} A plot showing the energies obtained using different single-code ans\"atze for various molecules. The solid brown line indicates the mean energy, while the error bars denote one standard deviation. The shaded region represents the range between the minimum and maximum energy values.} 
\end{figure*}

In Fig.~\ref{fig:SCAs}, we plot the energies corresponding to all the single-code ans\"atze for the various molecules investigated in this work.
For reference, we also include the energies of the Hartree–Fock and FCI states.

\section*{Function evaluations}
\begin{figure}[htbp!]
    \centering
    \includegraphics[width=0.5\textwidth]{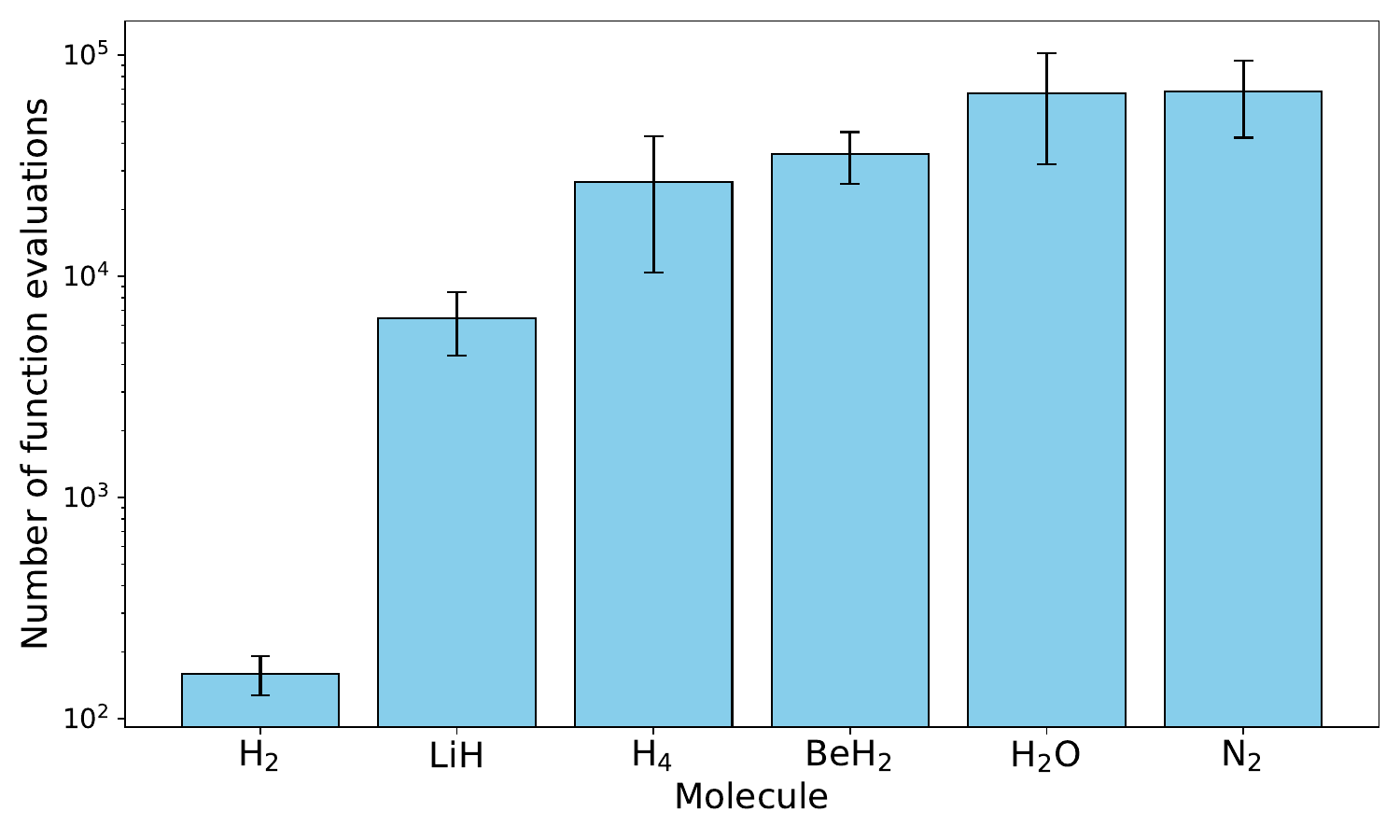}
    \caption{Average number of function evaluations required to compute the ground-state energy for different geometries of the molecules using a single-layer combined codes ansatz. Error bars indicate the standard deviation.}
    \label{fig:func_calls}
\end{figure}

In Fig.~\ref{fig:func_calls}, we present the number of function evaluations required to compute the ground-state energy using a single layer of the combined codes ansatz for each of the molecules studied in this work. 
The reported values represent averages over the different geometries considered for each molecule.

\end{document}